\documentclass[a4paper]{article}

\usepackage{latexsym}
\usepackage{a4wide}
\usepackage{graphicx, subfigure} 
\usepackage{cite} 
\usepackage{tabularx}
\usepackage{array}
\usepackage{graphics}     
\usepackage{amsmath}
\usepackage{amssymb}
\usepackage{longtable} 
\usepackage{verbatim} 
\usepackage[table]{xcolor}
\usepackage{booktabs}
\usepackage{hyperref} 
\usepackage[utf8]{inputenc}
\usepackage{multirow}
\usepackage{soul}

\definecolor{light-gray}{gray}{0.90}

\begin{document}

\hfill {\tt CERN-TH-2021-064, MITP/21-020}  

\def\thefootnote{\fnsymbol{footnote}}
 
\begin{center}

\vspace{3.cm}

{\Large\bf {More Indications for Lepton Nonuniversality in $b \to s \ell^+ \ell^-$}}

\setlength{\textwidth}{11cm}
                    
\vspace{2.cm}
{\large\bf  
T.~Hurth$^{a}$,\,
F.~Mahmoudi$^{b,c}$,\,
D.~Mart\'inez Santos$^{d}$,\,
S.~Neshatpour$^{b}$
}
 
\vspace{1.cm}
{\em $^a$PRISMA+ Cluster of Excellence and  Institute for Physics (THEP),\\
Johannes Gutenberg University, D-55099 Mainz, Germany}\\[0.2cm]
{\em $^b$Universit\'e de Lyon, Universit\'e Claude Bernard Lyon 1, CNRS/IN2P3, \\
  Institut de Physique des 2 Infinis de Lyon, UMR 5822, F-69622, Villeurbanne, France}\\[0.2cm]
{\em $^c$Theoretical Physics Department, CERN, CH-1211 Geneva 23, Switzerland}\\[0.2cm]
{\em $^d$Instituto Galego de F\'isica de Altas Enerx\'ias,\\ Universidade de Santiago de Compostela, Spain}\\[0.2cm]

\end{center}

\renewcommand{\thefootnote}{\arabic{footnote}}
\setcounter{footnote}{0}

\vspace{1.cm}
\thispagestyle{empty}
\centerline{\bf ABSTRACT}
\vspace{0.5cm}
Recently the LHCb collaboration has confirmed the evidence for lepton flavour nonuniversality at the $3.1\sigma$ level via an updated measurement of $R_K$. In this work we analyse this evidence within a model-independent approach. We make projections for future measurements which indicate that LHCb will be in the position to discover lepton nonuniversality with the Run 3 data in a single observable. We analyse other ratios based on our analysis of the present measurements of the ratios  $R_{K^{(*)}}$ and analyse if they are able to differentiate between various new physics options within the effective field theory at present or in the near future. We also compare the present  deviations in the ratios with NP indications in the angular observables of exclusive $b \to s \ell\ell$ transitions. Finally, we update our global analysis considering all $b \to s \ell\ell$ observables altogether, including a 20-parameter fit in connection of a Wilks' test. 

\newpage

\section{{Introduction}} 
Ever since the measurement of the full angular observables of the exclusive $B\to K^* \mu^+ \mu^-$ decay with 1~fb$^{-1}$ data by LHCb~\cite{Aaij:2013qta} which indicated New Physics (NP) in $C_9^\mu$~\cite{Descotes-Genon:2013wba,Altmannshofer:2013foa,Beaujean:2013soa,Horgan:2013pva,Hurth:2013ssa}, rare $b \to s$ observables have been showing the strongest hints for NP. Updated measurements of the $B\to K^* \mu^+ \mu^-$ angular observables by the LHCb experiment with 3 and 4.7 fb$^{-1}$ data~\cite{Aaij:2015oid,Aaij:2020nrf} as well as measurements in other exclusive modes such as $B_s \to \phi \mu^+ \mu^-$~\cite{Aaij:2015esa} have indicated signs of NP (with deviations of more than $2\sigma$ for some observables/bins). Although the SM predictions of some of the angular observables of the aforementioned modes have rather small uncertainties, in general the observables of the exclusive decays suffer from hadronic uncertainties, which often do not allow us to confidently separate possible NP effects from hadronic effects. 

Another group of rare decays which have shown signs of NP are lepton flavour universality violating (LFUV) observables  $R_{K^{(*)}}\equiv {\rm BR}(B^{+(0*)} \to K^{+(0*)} \mu^+ \mu^-)/{\rm BR}(B^{+(0*)} \to K^{+(0*)} e^+ e^-)$ ~\cite{Hiller:2003js} where the ratios of the branching fractions of muons compared to electrons are considered. The LFUV observables are theoretically very clean with SM uncertainties less than one percent\footnote{Only for the very low $q^2 \in [0.045, 1.1]$ GeV$^2$ bin of $R_{K^*}$ one finds an uncertainty of 
$\sim 3\%$~\cite{Bordone:2016gaq}.}.
The first tension in LFUV observables was measured for $R_K$ in the $[1.1,6.0]$ GeV$^2$ bin with the LHCb Run-1 data with $2.6\sigma$ significance~\cite{Aaij:2014ora}. This tension was confirmed with a signficance of $2.5\sigma$ when combining the Run 2 and the re-optimised Run 1 result~\cite{Aaij:2019wad} which had smaller uncertainty although the central value was measured to be closer to the SM prediction. LHCb found similar tensions at the level of $2.3$ and $2.5\sigma$ for $R_{K^*}$ in the two low-$q^2$ bins $[0.045, 1.1]$ and $[1.1,6.0]$~GeV$^2$, respectively~\cite{Aaij:2017vbb}. These tensions within the theoretically clean ratios  were shown to be rather consistent with the previously found tensions in the angular 
observables~\cite{Alonso:2014csa,Hiller:2014yaa,Hurth:2014vma,Altmannshofer:2014rta}.

Among the non-LFUV observables, the BR($B_s \to \mu^+ \mu^-$) is one of the cleanest observables giving a very good handle on the muon sector without involving the electron sector. Moreover, assuming no NP contributions due to scalar and pseudo-scalar operators (which is indicated by $b\to s \ell^+ \ell^-$ global fits), the short-distance contribution to this decay is only via $C_{10}^{\mu(\prime)}$.

Recently LHCb has updated two of the clean observables, namely $R_K$ and $BR(B_s \to \mu^+ \mu^-)$ using the complete dataset collected so far~\cite{LHCbtalkSantimaria,Aaij:2021vac}.
The LHCb experiment measures $3.1\sigma$ tension with the SM prediction for $R_K$
\begin{align}
R_K^{\rm LHCb}([1.1, 6.0]\;{\rm GeV}^2) = 0.846^{+0.042+0.013}_{-0.039-0.012}\,,
\end{align} 
which compared to the previous result with 5 fb$^{-1}$ data~\cite{Aaij:2019wad} has exactly the same central value but now due to smaller experimental uncertainties has an increased tension with the SM.

The new LHCb measurement of BR$(B_s \to \mu^+ \mu^-)$ gives~\cite{LHCbtalkSantimaria} 
\begin{align}
{\rm BR}(B_s \to \mu^+ \mu^-)^{\rm LHCb} = (3.09^{+0.46+0.15}_{-0.43-0.11})\times 10^{-9}\,.
\end{align}  

\begin{figure}[h!]
\begin{center}
\includegraphics[width=0.45\textwidth]{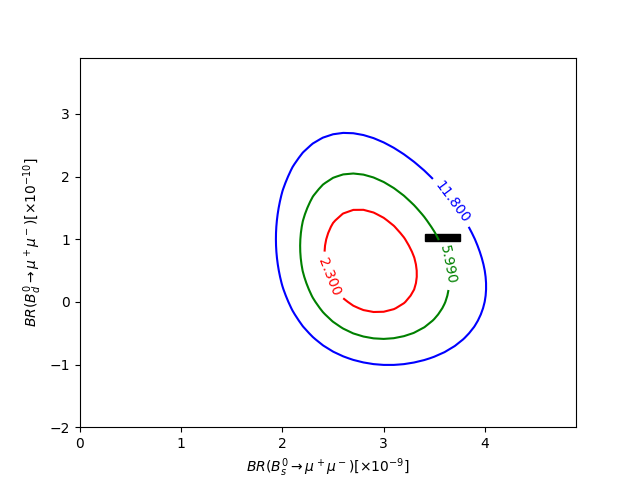}
\caption{\small 
2D likelihood plot of BR$(B_{s,d} \to \mu^+ \mu^-)$.
\label{fig:2Dlikelihood}}
\end{center}
\end{figure}

In our fits, for the experimental value of  BR($B_s \to \mu^+ \mu^-$) we combine the recent LHCb update~\cite{LHCbtalkSantimaria} with the ATLAS~\cite{Aaboud:2018mst} and CMS~\cite{Sirunyan:2019xdu} results considering a joint 2D likelihood as shown in Fig.~\ref{fig:2Dlikelihood}. For the combined experimental measurement of the $B_s \to \mu^+ \mu^-$ decay we have
\begin{align}\label{eq:Bsmumu_comb}
{\rm BR}(B_s \to \mu^+ \mu^-)^{\rm exp (comb.)} = (2.85_{-0.31}^{+0.34} )\times 10^{-9}\,.
\end{align}
\vspace{-1cm}

In the next section we discuss the impact of the recent LHCb measurements on the fit to clean observables and investigate in detail the role of BR($B_s \to \mu^+ \mu^-$) in the two-dimensional fit to LFUV observables. We also analyse the consistency of the clean observables and the rest of the $b \to s \ell \ell$  data regarding new physics.
In section~\ref{sec:NPglobal} we update our global $b \to s \ell \ell$ analysis in a multidimensional New Physics fit and consider the Wilks' test.
In section~\ref{sec:future}, we present the future prospects of the clean observables, and make predictions for further ratios. The conclusions are given in section~\ref{sec:conclusions}.

\section{{New physics analysis}}\label{sec:NPanalysis}
In this section we consider the new physics analysis of $R_{K^{(*)}}$ and BR($B_{s,d} \to \mu^+ \mu^-$) and compare with other $b \to s$ data.
The nonfactorisable power corrections in exclusive $b \to s \ell\ell$ decays are still not under control and have to be guesstimated, but promising approaches like the one in Ref.~\cite{Bobeth:2017vxj} may solve this problem in the near future (see Ref.~\cite{Gubernari:2020eft} for recent progress). Thus, it is still reasonable to make separate analyses of the theoretically very clean ratios and the other $b \to s$ observables to crosscheck the consistency of the two data sets.  

In Table~\ref{tab:ComparisonRKRKstarBmumu_1DA} the results of the one operator fits to new physics using only the data on $R_K$, $R_{K^*}$ and $B_{s,d} \to \mu\mu$ are shown where for our analysis we have used the {\ttfamily SuperIso} public program~\cite{Mahmoudi:2007vz}. 
Compared to our 2019 fits in Ref.~\cite{Arbey:2019duh} we see  increased significances for the NP fits to the theoretically clean ratios $R_K$ and $R_{K^*}$. In general, the SM pull of the one operator fits are changed by more than $1\sigma$. 
This is clearly due to the increased tension of the recent $R_K$ measurement with the SM. 
Our findings are in agreement with the recent model-independent analyses in Refs.~\cite{Geng:2021nhg,Altmannshofer:2021qrr,Alguero:2021anc,Angelescu:2021lln}.  
\begin{table}[th!]
\begin{center}
\setlength\extrarowheight{0pt}
\scalebox{0.85}{
{
\begin{tabular}{|l|r|r|c|}
\hline 
 \multicolumn{4}{|c|}{\footnotesize Only $R_K, R_{K^*}, B_{s,d}\to \mu^+ \mu^-$ \vspace{-0.1cm}} \\
 \multicolumn{4}{|c|}{\small{\bf 2019 data} \; ($\chi^2_{\rm SM}=19.0$)} \\ \hline
                          & b.f. value & $\chi^2_{\rm min}$ & ${\rm Pull}_{\rm SM}$  \\ 
\hline \hline
$\delta C_{9} $          		& $ 	-2.04	\pm	5.93	 $ & $	18.9	 $ & $ 	0.3	\sigma	 $  \\
$\delta C_{9}^{e} $        		& $ 	0.79	\pm	0.29	 $ & $	9.9	 $ & $ 	3.0	\sigma	 $  \\
$\delta C_{9}^{\mu} $      		& $ 	-0.74	\pm	0.28	 $ & $	10.6	 $ & $ 	2.9	\sigma	 $  \\
\hline
$\delta C_{10} $         		& $ 	0.43	\pm	0.32	 $ & $	17.0	 $ & $ 	1.4	\sigma	 $  \\
$\delta C_{10}^{e} $       		& $ 	-0.78	\pm	0.27	 $ & $	8.2	 $ & $ 	3.3	\sigma	 $  \\
$\delta C_{10}^{\mu} $     		& $ 	0.65	\pm	0.20	 $ & $	6.9	 $ & $ 	3.5	\sigma	 $  \\
\hline
$\delta C_{\rm LL}^e$	& $ 	0.41	\pm	0.15	 $ & $	9.0	 $ & $ 	3.2	\sigma	 $  \\
$\delta C_{\rm LL}^\mu$	& $ 	-0.37	\pm	0.11	 $ & $	7.2	 $ & $ 	3.4	\sigma	 $  \\
\hline
\end{tabular}
}\qquad
\begin{tabular}{|l|r|r|c|}
\hline 
\multicolumn{4}{|c|}{\footnotesize Only $R_{K^{(*)}}, B_{s,d} \to \mu^+ \mu^-$ \vspace{-0.1cm}}\\	 									
\multicolumn{4}{|c|}{\small{\bf 2021 data} \; ($\chi^2_{\rm SM}=	28.19	$)}\\ \hline								
& b.f. value & $\chi^2_{\rm min}$ & ${\rm Pull}_{\rm SM}$  \\										
\hline \hline										
$\delta C_{9} $    	& $ 	-1.00	\pm	6.00	 $ & $ 	28.1	 $ & $	0.2	\sigma	 $  \\
$\delta C_{9}^{e} $    	& $ 	0.80	\pm	0.21	 $ & $ 	11.2	 $ & $	4.1	\sigma	 $  \\
$\delta C_{9}^{\mu} $    	& $ 	-0.77	\pm	0.21	 $ & $ 	11.9	 $ & $	4.0	\sigma	 $  \\
\hline										
$\delta C_{10} $    	& $ 	0.43	\pm	0.24	 $ & $ 	24.6	 $ & $	1.9	\sigma	 $  \\
$\delta C_{10}^{e} $    	& $ 	-0.78	\pm	0.20	 $ & $ 	9.5	 $ & $	4.3	\sigma	 $  \\
$\delta C_{10}^{\mu} $    	& $ 	0.64	\pm	0.15	 $ & $ 	7.3	 $ & $	4.6	\sigma	 $  \\
\hline							          			
$\delta C_{\rm LL}^e$	& $ 	0.41	\pm	0.11	 $ & $ 	10.3	 $ & $	4.2	\sigma	 $  \\
$\delta C_{\rm LL}^\mu$	& $ 	-0.38	\pm	0.09	 $ & $ 	7.1	 $ & $	4.6	\sigma	 $  \\
\hline										
\end{tabular}
} 
\caption{Comparison of one operator NP fits with 2019 and 2021 
data on $R_K, R_{K^*}$ and BR($B_{s,d}\to \mu^+ \mu^-$). 
\label{tab:ComparisonRKRKstarBmumu_1DA}} 
\end{center} 
\end{table}
\vspace{-0.2cm}

For the rest of the $b \to s \ell^+ \ell^-$ observables (except  $R_K$, $R_{K^*}$ and $B_{s,d} \to \mu\mu$), we also find larger SM pulls of the one operator fits compared to the analysis in Ref~\cite{Arbey:2019duh}. As shown in Table~\ref{tab:ComparisonRKRKstarBmumu_1DB}, there is a $1-2\sigma$ increase of the SM pull which is due to a new measurement of the angular observables of the $B^{0} \to K^{*0}\mu^+\mu^-$ decay (see Ref~\cite{Hurth:2020rzx} for more details)  and due to the inclusion of further observables such as the angular observables of the charged $B^{+} \to K^{*+}\mu^+\mu^-$ decay (see Ref.~\cite{Hurth:2020ehu} ) and the branching ratio and angular observables of the $\Lambda_b \to \Lambda \mu^+\mu^-$ baryonic decays\footnote{The complete list of $b \to s \ell^+ \ell^-$ observables are as given in Ref.~\cite{Hurth:2020rzx} together with angular observables of the $B^{+} \to K^{*+}\mu^+\mu^-$ decay.}.
We emphasize again that the large SM-pulls beyond $5\sigma$ are based on guesstimates of the 10\% nonfactorisable power corrections within the angular observables.

\begin{table}[h!]
\begin{center}
\setlength\extrarowheight{0pt}
\scalebox{0.85}{
{
\begin{tabular}{|l|r|r|c|}
\hline 
 \multicolumn{4}{|c|}{\footnotesize All observables except $R_{K^{(*)}}, B_{s,d}\to \mu^+ \mu^-$ \vspace{-0.1cm}} \\ 
 \multicolumn{4}{|c|}{\small{\bf 2019 data} \;($\chi^2_{\rm SM}=99.7$)} \\ \hline
                          & b.f. value & $\chi^2_{\rm min}$ & ${\rm Pull}_{\rm SM}$  \\ 
\hline \hline
$\delta C_{9} $          	& $ 	-1.03	\pm	0.20	 $ & $	81.0	 $ & $ 	4.3	\sigma	 $  \\
$\delta C_{9}^{e} $        	& $ 	0.72	\pm	0.58	 $ & $	98.5	 $ & $ 	1.1	\sigma	 $  \\
$\delta C_{9}^{\mu} $      	& $ 	-1.05	\pm	0.19	 $ & $	78.8	 $ & $ 	4.6	\sigma	 $  \\
\hline
$\delta C_{10} $         	& $ 	0.27	\pm	0.28	 $ & $	98.7	 $ & $ 	1.0	\sigma	 $  \\
$\delta C_{10}^{e} $       	& $ 	-0.56	\pm	0.50	 $ & $	98.7	 $ & $ 	1.0	\sigma	 $  \\
$\delta C_{10}^{\mu} $     	& $ 	0.38	\pm	0.28	 $ & $	97.7	 $ & $ 	1.4	\sigma	 $  \\
\hline
$\delta C_{\rm LL}^e$	& $ 	0.33	\pm	0.29	 $ & $	98.6	 $ & $ 	1.1	\sigma	 $  \\
$\delta C_{\rm LL}^\mu$	& $ 	-0.50	\pm	0.16	 $ & $	88.8	 $ & $ 	3.3	\sigma	 $  \\
\hline
\end{tabular}
}\qquad
\begin{tabular}{|l|r|r|c|}
\hline 
\multicolumn{4}{|c|}{\footnotesize All observables except $R_{K^{(*)}}, B_{s,d}\to\mu^+ \mu^-$ \vspace{-0.1cm}}\\	 									
\multicolumn{4}{|c|}{\small{\bf 2021 data} \; ($\chi^2_{\rm SM}=	200.1	$)}\\ \hline								
& b.f. value & $\chi^2_{\rm min}$ & ${\rm Pull}_{\rm SM}$  \\	 									
\hline \hline										
$\delta C_{9} $    	& $ 	-1.01	\pm	0.13	 $ & $ 	158.2	 $ & $	6.5	\sigma	 $  \\
$\delta C_{9}^{e} $    	& $ 	0.70	\pm	0.60	 $ & $ 	198.8	 $ & $	1.1	\sigma	 $  \\
$\delta C_{9}^{\mu} $    	& $ 	-1.03	\pm	0.13	 $ & $ 	156.0	 $ & $	6.6	\sigma	 $  \\
\hline										
$\delta C_{10} $    	& $ 	0.34	\pm	0.23	 $ & $ 	197.7	 $ & $	1.5	\sigma	 $  \\
$\delta C_{10}^{e} $    	& $ 	-0.50	\pm	0.50	 $ & $ 	199.0	 $ & $	1.0	\sigma	 $  \\
$\delta C_{10}^{\mu} $    	& $ 	0.41	\pm	0.23	 $ & $ 	196.5	 $ & $	1.9	\sigma	 $  \\
\hline							          			
$\delta C_{\rm LL}^e$	& $ 	0.33	\pm	0.29	 $ & $ 	198.9	 $ & $	1.1	\sigma	 $  \\
$\delta C_{\rm LL}^\mu$	& $ 	-0.75	\pm	0.13	 $ & $ 	167.9	 $ & $	5.7	\sigma	 $  \\
\hline										
\end{tabular}\qquad
} 
\caption{\small Comparison of one operator NP fits with 2019 and 2021 data on $b \to s \ell^+ \ell^-$ transitions except $R_K, R_{K^*}$, BR($B_{s,d}\to \mu^+ \mu^-$) assuming 10\% error  for the power corrections.
The observable set used in 2019 fit on the left panel is not exactly the same as the 2021 one (e.g. $B^+\to K^{*+} \mu^+ \mu^-$ and $\Lambda_b \to \Lambda \mu^+\mu^-$ observables).
\label{tab:ComparisonRKRKstarBmumu_1DB}} 
\end{center} 
\end{table}

\begin{figure}[h!]
\begin{center}
\includegraphics[width=0.49\textwidth]{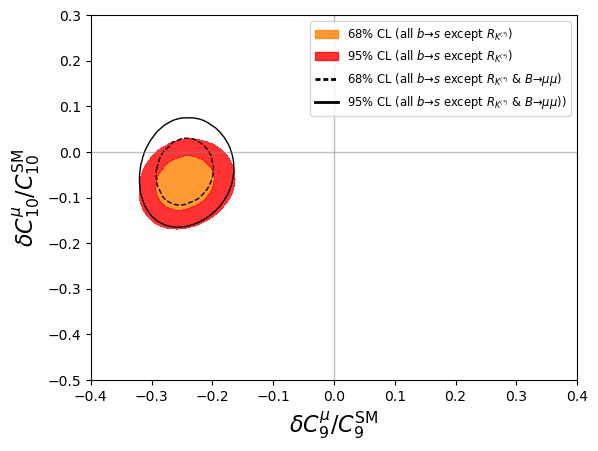}
\includegraphics[width=0.49\textwidth]{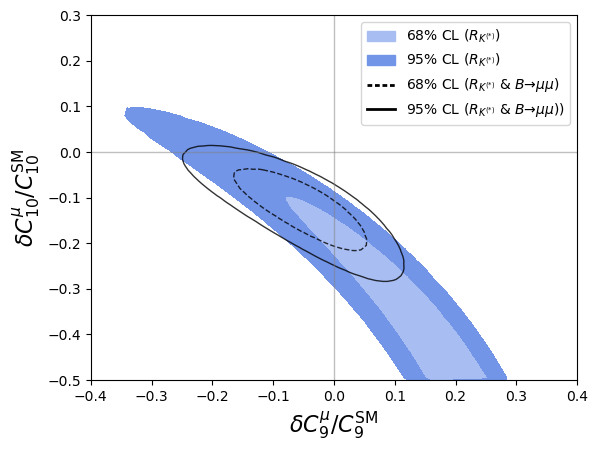}
\caption{\small {\color{red}}  
Two operator fits to $\{C_9^\mu, C_{10}^\mu\}$. 
On  the left  we have considered all observables except $R_K$ and $R_{K^*}$ with the assumption of 10\% power corrections giving Pull$_{\rm SM} = 6.4\sigma$.
On the right we have only used the data on $R_K, R_{K^*}$ finding Pull$_{\rm SM} = 4.1\sigma$. 
The black  dashed and solid contours correspond to excluding (including) the data on $B_{s,d}\to \mu^+ \mu^-$ from (to) the fits of the left (right) plot where Pull$_{\rm SM}$ remains the same  (becomes $4.2\sigma$).
\label{fig:RKRKstarBmumu}}
\end{center}
\end{figure}

We also present two operator fits and analyse the role of the modes $B_{s,d}\to \mu^+ \mu^-$:   
The one-dimensional fits are coherent in the preferred NP scenario whether the clean observables are considered for the fit, or the rest of the $b\to s \ell \ell$ data, indicating either a negative $\delta C_9$ or a positive $\delta C_{10}$ for both sets of observables. 
This coherence is however not trivial in the two operator fits. In the fit to the clean observables it is crucial to also consider BR($B_{s,d}\to \mu^+\mu^-$) in order to get the correct sign for $\{C_{9},C_{10}\}$ as shown in Fig.~\ref{fig:RKRKstarBmumu} where the best fit value of the fit to only LFUV ratios (the colored region of the right plot) indicates positive $\delta C_{10}$ as well as a positive $\delta C_9$, and it is only after  including   BR($B_{s,d}\to \mu^+\mu^-$) (the black contour of the right plot) that similar best fit signs are obtained for the clean and the rest of the observables.
%
\begin{figure}[h!]
\begin{center}
\includegraphics[width=0.49\textwidth]{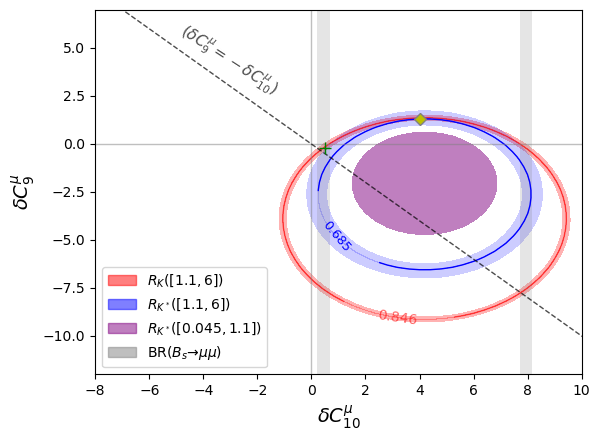}
\includegraphics[width=0.49\textwidth]{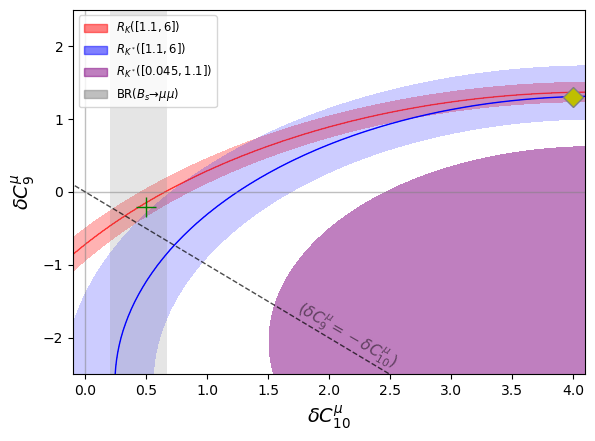}
\caption{\small{
Clean observables within 1$\sigma$ of experimental central value.
The red (blue) solid line indicate the central value of the experimental measurements of 
$R_K^{[1.1,6]}$ ($R_{K^*}^{[1.1,6]}$). The colored regions give the $1\sigma$ range
(theoretical and experimental uncertainties added in quadrature) with the experimental central value.  
The yellow diamond corresponds to the best fit point  
($\{C_9^\mu,C_{10}^\mu \} = \{1.3\pm 0.1, 4.0\pm4.0 \}$) 
of the fit to  only $R_{K^{(*)}}$ 
while the green cross indicates the best fit value 
($\{C_9^\mu,C_{10}^\mu \} = \{-0.2\pm 0.3, 0.5\pm0.2 \}$)
when fitting  to $R_{K^{(*)}}$ and BR($B_{s,d} \to \mu^+ \mu^-$).}
\label{fig:C9C10degeneracy}}
\end{center}
\end{figure}

This feature is due to the degeneracy that the ratios $R_{K^{(*)}}$ have in $C_{9}^\mu$ and $C_{10}^\mu$ as can be seen by the circular contours of Fig.~\ref{fig:C9C10degeneracy} where a positive $\delta C_9^\mu$ explains the data when simultaneously having a rather large value of $\delta C_{10}^\mu$ (not consistent with other $b \to s \ell\ell$ observables). 
The best fit value of the fit to only $R_{K^{(*)}}$ is $\{C_9^\mu,C_{10}^\mu \} = \{1.3\pm 0.1, 4.0\pm4.0 \}$ as indicated with the yellow diamond in Fig.~\ref{fig:C9C10degeneracy} while when the BR($B_{s,d}\to \mu^+\mu^-$) is also included the best fit value is $\{C_9^\mu,C_{10}^\mu \} = \{-0.2\pm 0.3, 0.5\pm0.2 \}$ as indicated with a green cross.

It should be noted that the LHCb measured central value of $R_{K^*}$ in the very low $q^2$ bin cannot be reached with any combination of NP in $\delta C_{9,10}^\mu$ since this bin is dominated by the photon contribution via the radiative Wilson coefficient $C_7$.

The comparison of the fits to the two separate sets of observables confirm our observations in previous analyses. Within the one operator fits the $C_{10}$-like Wilson coefficients  play a significant role in the set of clean observables, {but} not in the complementary set. And the two operator fits in Figure\,\ref{fig:RKRKstarBmumu} indicate that there is consistency between the two sets of observables at the 2$\sigma$ level only.

\section{{Global fit}}\label{sec:NPglobal} 
\begin{figure}[h!]
\begin{center}
\includegraphics[width=0.33\textwidth]{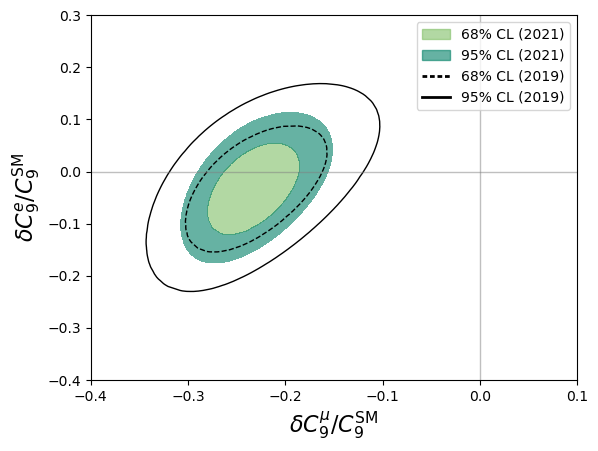}
\includegraphics[width=0.32\textwidth]{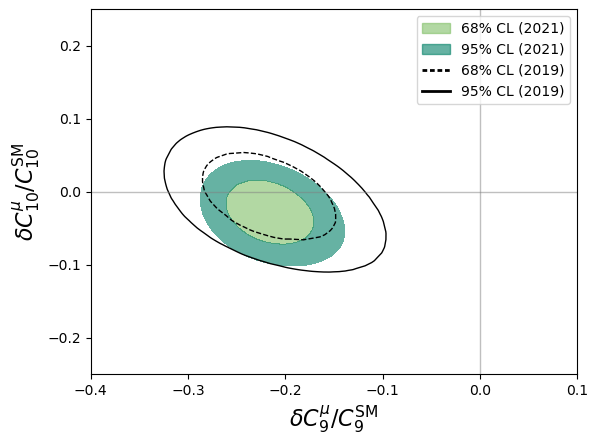}
\includegraphics[width=0.32\textwidth]{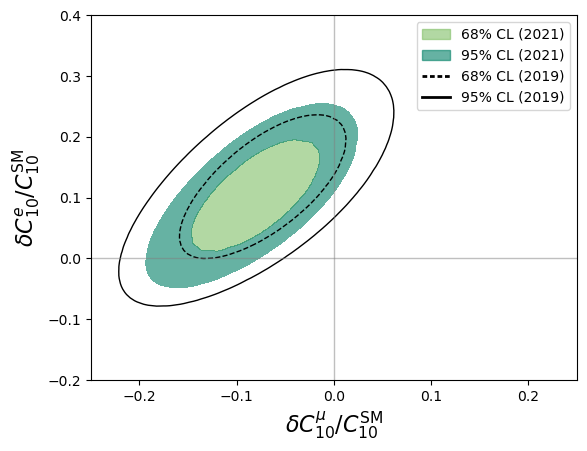}
\\[2.mm]
\includegraphics[width=0.32\textwidth]{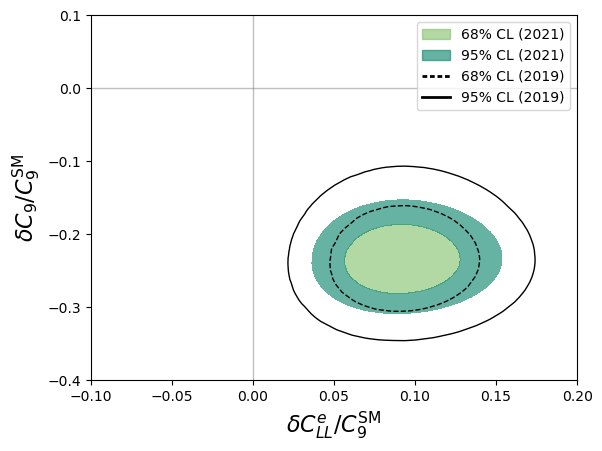}\qquad
\includegraphics[width=0.32\textwidth]{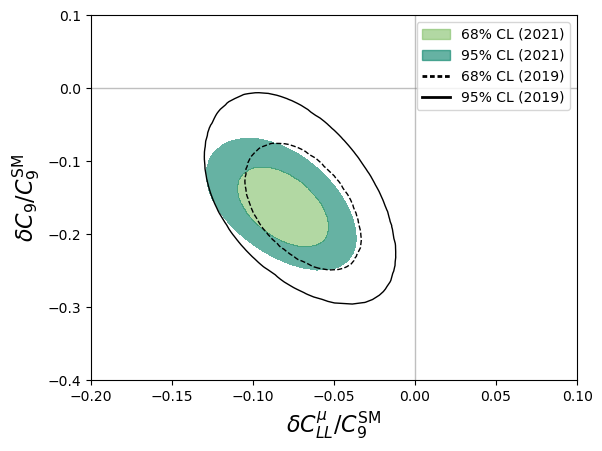}
\caption{\small Two operator fits to NP, considering all observables (with the assumption of 10\% power corrections). The colored bands (black contours) correspond to the 68 and 95\% confidence level regions considering the 2021 (2019) data.
Pull$_{\rm SM}$ in the $\{\delta C_{9}^e , \delta C_{9}^\mu\}, \{\delta C_{10}^\mu , \delta C_{9}^\mu\}, \{\delta C_{10}^{e} , \delta C_{10}^{\mu}\}$ fits are $7.3, 7.3, 4.5\sigma$, respectively. 
Pull$_{\rm SM}$ for the $\{\delta C_{LL}^{e},\delta C_9\}$ and $\{\delta C_{LL}^{\mu},\delta C_9\}$ fits of the lower row are $7.4$ and $7.5\sigma$, respectively.
\label{fig:all}}
\end{center}
\end{figure}

In the next step we show the global one and two operator fits in Table~\ref{tab:ALL_1D} and in Fig.~\ref{fig:all} respectively using the $b \to s$ data altogether. 
In the one operator fit, the hierarchy of the preferred NP scenarios have remained the same as is in 2019, with the most prominent scenario indicating beyond the SM contributions to the muon Wilson coefficient $\delta C_9^\mu$ followed by $\delta C_{LL}^\mu$ and the universal (not lepton flavour-dependent) Wilson coefficient $\delta C_9$. 
The significance of these scenarios have increased by more than $2\sigma$ compared to 2019 using the same 10\% assumption for the power corrections. This increase is mainly due to the updated measurement of the $B^{0} \to K^{*0} \mu^+ \mu^-$ angular observables as well as the measurement of $B^{+} \to K^{*+} \mu^+ \mu^-$ and finally  the recent LHCb measurements of $R_K$ and $B_s \to \mu^+\mu^-$ where interestingly at each step the data has indicated the same preferred NP scenario (see also Refs.\cite{Hurth:2020ehu,Hurth:2020rzx}). 
\begin{table}[th!]
\begin{center}
\setlength\extrarowheight{0pt}
\scalebox{0.85}{
\begin{tabular}{|l|r|r|c|}
\hline 
\multicolumn{4}{|c|}{All observables  \vspace{-0.1cm}}\\	 									
\multicolumn{4}{|c|}{\small{\bf 2019 data} \; ($\chi^2_{\rm SM}=	117.03	$)}\\ \hline		& b.f. value & $\chi^2_{\rm min}$ & ${\rm Pull}_{\rm SM}$  \\	
\hline \hline
$\delta C_{9} $          	& $ 	-1.01	\pm	0.20	 $ & $ 	99.2	 $ & $	4.2	\sigma	 $  \\
$\delta C_{9}^{e} $        	& $ 	0.78	\pm	0.26	 $ & $ 	106.6	 $ & $	3.2	\sigma	 $  \\
$\delta C_{9}^{\mu} $      	& $ 	-0.93	\pm	0.17	 $ & $ 	89.4	 $ & $	5.3	\sigma	 $  \\
\hline
$\delta C_{10} $         	& $ 	0.25	\pm	0.23	 $ & $ 	115.7	 $ & $	1.1	\sigma	 $  \\
$\delta C_{10}^{e} $       	& $ 	-0.73	\pm	0.23	 $ & $ 	105.2	 $ & $	3.4	\sigma	 $  \\
$\delta C_{10}^{\mu} $     	& $ 	0.53	\pm	0.17	 $ & $ 	105.8	 $ & $	3.3	\sigma	 $  \\
\hline
$\delta C_{\rm LL}^e$		& $ 	0.40	\pm	0.13	 $ & $ 	105.8	 $ & $	3.3	\sigma	 $  \\
$\delta C_{\rm LL}^\mu$ 	& $ 	-0.41	\pm	0.10	 $ & $ 	96.6	 $ & $	4.5	\sigma	 $  \\
\hline
\end{tabular}\qquad
\begin{tabular}{|l|r|r|c|}
\hline 
\multicolumn{4}{|c|}{All observables  \vspace{-0.1cm}}\\	 									
\multicolumn{4}{|c|}{\small{\bf 2021 data} \; ($\chi^2_{\rm SM}=	225.8	$)}\\ \hline		& b.f. value & $\chi^2_{\rm min}$ & ${\rm Pull}_{\rm SM}$  \\													
\hline \hline										
$\delta C_{9} $    	& $ 	-0.99	\pm	0.13	 $ & $ 	186.2	 $ & $	6.3	\sigma	 $  \\
$\delta C_{9}^{e} $    	& $ 	0.79	\pm	0.20	 $ & $ 	207.7	 $ & $	4.3	\sigma	 $  \\
$\delta C_{9}^{\mu} $    	& $ 	-0.95	\pm	0.12	 $ & $ 	168.6	 $ & $	7.6	\sigma	 $  \\
\hline										
$\delta C_{10} $    	& $ 	0.32	\pm	0.18	 $ & $ 	222.3	 $ & $	1.9	\sigma	 $  \\
$\delta C_{10}^{e} $    	& $ 	-0.74	\pm	0.18	 $ & $ 	206.3	 $ & $	4.4	\sigma	 $  \\
$\delta C_{10}^{\mu} $    	& $ 	0.55	\pm	0.13	 $ & $ 	205.2	 $ & $	4.5	\sigma	 $  \\
\hline							          			
$\delta C_{\rm LL}^e$	& $ 	0.40	\pm	0.10	 $ & $ 	206.9	 $ & $	4.3	\sigma	 $  \\
$\delta C_{\rm LL}^\mu$	& $ 	-0.49	\pm	0.08	 $ & $ 	180.5	 $ & $	6.7	\sigma	 $  \\
\hline										
\end{tabular}\qquad
}
\caption{\small
Comparison of one operator NP fits with 2019 and 2021 data on all the relevant data on $b \to s$ transitions, assuming 10\% error for the power corrections. 
\label{tab:ALL_1D}} 
\end{center} 
\end{table}

For the two operator fit the most prominent scenario still involves a universal NP contribution to  $\delta C_9$ together with $\delta C_{LL}^\mu$ very slightly preferred over $\delta C_{LL}^e$ followed by NP in the 
$\{\delta C_{9}^e , \delta C_{9}^\mu\}$ and $\{\delta C_{10}^\mu , \delta C_{9}^\mu\}$ as given in the caption of Fig.~\ref{fig:all}.
All the four mentioned scenarios have very similar Pull$_{\rm SM}$ and all involve a NP contribution to $\delta C_9^{(\mu)}$.  
Our results are in part  consistent with, but also in part different from the results in the recent model-independent analyses in Refs.~\cite{Geng:2021nhg,Altmannshofer:2021qrr,Alguero:2021anc,Alok:2019ufo} (see also~\cite{Cacciapaglia:2021gff}). There are two obvious reasons which are responsible for larger discrepancies in the SM-pulls, namely  different guesstimates of the nonfactorisable power corrections and different choices of the set of observables used in the fit.

In general NP contributions do not necessarily contribute to only one or two Wilson coefficients and could simultaneously involve several operator structures. In these cases the one and two operator fits lead to unnaturally large SM-pulls. 
Indeed, beyond simplified models, general NP scenarios contain a variety of new particles and new couplings.
Therefore, taking a more agnostic approach to the behaviour of NP contributions, as first proposed in Refs.~\cite{Arbey:2018ics,Mahmoudi:2018qsk,Hurth:2018kcq} we make a 20-dimensional fit, varying all the relevant $b \to s $ Wilson coefficients, thus, considering the most general description of NP effects in the $b \to s\ell\ell$ channel. We also established criteria to identify possible insensitive parameters and flat directions regarding NP.   We note that our method avoids any look-elsewhere effect by starting with the most general description of possible NP effects and by eliminating insensitive parameters and flat directions based on the fit and not based on data.
For an alternative approach to include the look-elsewhere effect see Ref.~\cite{Lancierini:2021sdf}.

\begin{table}[h!]
\begin{center}
\setlength\extrarowheight{3pt}
\scalebox{0.85}{
\begin{tabular}{|c|c|c|c|}
\hline																
\multicolumn{4}{|c|}{All observables  with $\chi^2_{\rm SM}=	 	225.8			$} \\											
\multicolumn{4}{|c|}{$\chi^2_{\rm min}=	 	151.6	;\quad {\rm Pull}_{\rm SM}=	5.5 (5.6)	\sigma$} \\											
\hline \hline																
\multicolumn{2}{|c|}{$\delta C_7$} &  \multicolumn{2}{c|}{$\delta C_8$}\\																
\multicolumn{2}{|c|}{$	0.05	\pm	0.03	$} & \multicolumn{2}{c|}{$	-0.70	\pm	0.40	$}\\								
\hline																
\multicolumn{2}{|c|}{$\delta C_7^\prime$} &  \multicolumn{2}{c|}{$\delta C_8^\prime$}\\																
\multicolumn{2}{|c|}{$	-0.01	\pm	0.02	$} & \multicolumn{2}{c|}{$	0.00	\pm	0.80	$}\\								
\hline																
$\delta C_{9}^{\mu}$ & $\delta C_{9}^{e}$ & $\delta C_{10}^{\mu}$ & $\delta C_{10}^{e}$ \\																
$	-1.16	\pm	0.17	$ & $	-6.70	\pm	1.20	$ & $	0.20	\pm	0.21	$ & degenerate w/ $C_{10}^{\prime e}$ \\
\hline\hline																
$\delta C_{9}^{\prime \mu}$ & $\delta C_{9}^{\prime e}$ & $\delta C_{10}^{\prime \mu}$ & $\delta C_{10}^{\prime e}$ \\																
$	0.09	\pm	0.34	$ & $	1.90	\pm	1.50	$ & $	-0.12	\pm	0.20	$ & degenerate w/ $C_{10}^{ e}$ \\
\hline\hline																
$C_{Q_{1}}^{\mu}$ & $C_{Q_{1}}^{e}$ & $C_{Q_{2}}^{\mu}$ & $C_{Q_{2}}^{e}$ \\																
$	0.04	\pm	0.10	$ & $	-1.50	\pm	1.50	$ & $	-0.09	\pm	0.10	$ & $	-4.10	\pm	1.5	$ \\[-6pt]
[$	-0.08	\pm	0.11	$] & [$	-0.20	\pm	1.60	$] & [$	-0.11	\pm	0.10	$] & [$	4.50	\pm	1.5	$] \\
\hline\hline																
$C_{Q_{1}}^{\prime \mu}$ & $C_{Q_{1}}^{\prime e}$ & $C_{Q_{2}}^{\prime \mu}$ & $C_{Q_{2}}^{\prime e}$ \\																
$	0.15	\pm	0.10	$ & $	-1.70	\pm	1.20	$ & $	-0.14	\pm	0.11	$ & $	-4.20	\pm	1.2	$ \\[-6pt]
[$	0.02	\pm	0.12	$] & [$	-0.30	\pm	1.10	$] & [$	-0.16	\pm	0.10	$] & [$	4.40	\pm	1.2	$] \\
\hline																
\end{tabular}
} 
\caption{\small Best fit values for the 20 operator global fit to the $b \to s$ data, assuming 10\% error for the power corrections.
The Pull$_{\rm SM}$ in the parenthesis corresponds considering 19 effective d.o.f. instead of 20.  
The numbers in the brackets refer to an alternative solution giving an equally good fit. 
\label{tab:ALL_20D_C78910C12primes}} 
\end{center} 
\end{table}

The results in Table~\ref{tab:ALL_20D_C78910C12primes} show that compared to our  previous analyses \cite{Arbey:2018ics,Arbey:2019duh}  we find  now that the fit constrains all four  parameters $C_{Q_1}^e,C_{Q_2}^e.C_{Q_1}^{e'}, C_{Q_2}^{e'}$ which {previously} were shown to be undetermined due to their large uncertainties in the previous analyses and negligible impact on the fit (resulting in the number of effective  {degrees of freedom to be 16). In Ref.\cite{Arbey:2018ics} a criterium was presented to single out  such undetermined parameters~\footnote{The precise criterium was given by the explicit check if the variation of each coefficient of order one, $|C_i| \sim 1$, implies/leads to  $|\delta \chi^2|<1$.}.

But also in the present fit the parameters $C_{10}^e$ and $C_{10}^{e'}$ have larger uncertainties and can be shown to have a small impact on the fit. In addition one finds that there is degeneracy in the sense that the fit constrains the difference of these two WCs only,  so for both WCs large values are possible. Removing one of the two WCs from the fit one finds the other one is well-constrained and does not  effectively change the $\chi^2$.   Therefore we have 19 effective degrees of freedom. 

The Pull$_{\rm SM}$ of the 20-parameter fit has increased by more than $2\sigma$ compared to the 2019 results. If we consider the fit with the 19 effective parameters we currently find a SM pull of $5.6\sigma$.

The Wilks' test allows us to estimate the impact of the various parameters even further. The likelihood ratio test via Wilks' theorem  enables us to estimate the significance of adding Wilson coefficients into a fit when one goes from one nested scenario to a more general one. In our previous analysis in 
Ref.\cite{Arbey:2018ics} we found that adding Wilson coefficients to the ``$C_9^\mu$ only'' scenario was improving the fit only marginally. The Wilks' test with the present data shows that adding $C_ {10}^\mu$ and $C_ {9}^e,C_ {10}^e$ and also $C_7$ and $C_8$ improves the fit significantly and establishes the importance of these fit parameters (see Table~\ref{tab:Wilks}).   This can be explained by the fact that to a great degree, the tension and its increase compared to our previous analysis in Ref.~\cite{Arbey:2018ics} is due to the updated LFUV ratio $R_K$ which can be described  equally well by NP contributions to the electron and muon sectors. Furthermore, there is now more data on observables with electrons in the final state.

\begin{table}[!t]
\begin{center}
\scalebox{0.95}{
\begin{tabular}{|c|c|c|c|c|}\hline
Set of WC & param. & $\chi^2_{\rm min}$ & Pull$_{\rm SM}$ & Improvement\\
\hline
SM                                          & 0   & 225.8 & - & -\\
$C_9^{\mu}$                                 & 1   & 168.6 & $7.6\sigma$ & $7.6\sigma$\\
$C_9^{\mu}, C_{10}^{\mu}$                   & 2   & 167.5 & $7.3\sigma$ & $1.0\sigma$\\
$C_7,C_8,C_9^{(e,\mu)},C_{10}^{(e,\mu)}$    & 6   & 158.0 & $7.1\sigma$ & $2.0\sigma$\\
All non-primed WC                           & 10  & 157.2 & $6.5\sigma$ & $0.1\sigma$\\
All WC (incl. primed)                       & 20 (19)  & 151.6 & $5.5\,(5.6)\sigma$ & $0.2\,(0.3)\sigma$\\
\hline
\end{tabular}
}
\caption{\small Pull$_{\rm SM}$ of $1,2,6,10$ and 20 dimensional fit. 
The ``All non-primed WC'' includes in addition to the previous row, the scalar and pseudoscalar Wilson coefficients. The last row also includes the chirality-flipped counterparts of the Wilson coefficients.
In the last column the significance of improvement of the fit compared to 
the scenario of the previous row is given. The number in parentheses corresponds 
to the effective degrees of freedom (see the text for further details).
\label{tab:Wilks}}\vspace*{-0.6cm}
\end{center} 
\end{table}

\section{Future prospects and predictions for other ratios} \label{sec:future} Upgrades of the LHCb experiment are planned. The first upgrade will lead to a total integrated luminosity of 50 fb$^{-1}$. A second upgrade at a high-luminosity LHC will lead to an integrated luminosity of 300 fb$^{-1}$. We also use a third intermediate benchmark with an integrated luminosity of 18 fb$^{-1}$ to analyse the future prospects of the clean observables $R_K$, $R_{K^*}$ and BR($B_s \to \mu^+ \mu^-$). 

Our estimates of the future systematical uncertainties are based on the following considerations.
From Table 2 in~\cite{Aaij:2021vac}, the efficiency ratio between the electron mode and the muon mode is approximately one-third. The LHCb Upgrade will replace the hardware trigger by a software trigger which  is expected to yield electron efficiencies closer to muon efficiencies (see Table 2 in~\cite{Aaij:2019zbu}). We assume the efficiency ratio in the LHCb Upgrade grows from one-third to~$\sim60\%$.
The ultimate systematic uncertainty for $R_K$ is expected to be~$\sim 1\%$ ~\cite{LHCbtalkPetridis}, and we assume that a similar ultimate systematic could hence be achieved for $R_{K^*}$ (also in line with ~\cite{Aaij:2244311}).
For the $B_s^0\rightarrow\mu^+\mu^-$ branching fraction, two important systematic sources depend on external information: the value of the $b$-quark hadronisation fractions $f_d/f_s$, and the branching fraction of the $B^+\rightarrow J/\psi(\rightarrow\mu^+\mu^-) K^+$ decay. Hence, an irreducible systematic of~$\sim4\%$ is assumed in our approach. We also assume that ATLAS, CMS, and LHCb will keep relative weights in the $B_s^0\rightarrow\mu^+\mu^-$ branching fraction world average similar to the ones they currently have.

Considering the decrease in experimental uncertainties as described above we investigate the NP fits to the clean observables. 
Keeping the experimental central values as what they are currently do not give acceptable fits which is partly due to the different preferred NP scenarios for $R_K$ and $R_{K^{*}}$ in the [1.1, 6] GeV$^2$ bin and partly due to the rather small value of $R_{K^{*}}$ in the [0.045, 1.1] GeV$^2$ bin which cannot be reached with NP in the preferred NP scenarios. Instead we make a different but similarly strong assumption that future experimental results are in agreement with one of the current NP scenarios from the fit to clean observables.

Considering the three preferred scenarios of Table~\ref{tab:ComparisonRKRKstarBmumu_1DA} we make projections for  18, 50 and 300~fb$^{-1}$ luminosities. For the 18~fb$^{-1}$ benchmark we assume 1.5, 3, 5 and  4.4\% systematic uncertainties for $R_K([1.1,6])$, $R_{K^*}([0.045,1.1])$, $R_{K^*}([1.1,6])$ and BR($B_s \to \mu^+ \mu^-$), respectively, while for the 50 and 300~fb$^{-1}$ projections we consider the ultimate systematic uncertainty of 1\% for all three LFUV ratios and 4\% for BR($B_s \to \mu^+ \mu^-$).
In all three scenarios the NP significance will be larger than $6\sigma$ as given in Table~\ref{tab:upgrade+Bsmumu}.

\begin{table}[h!]
\begin{center}
\setlength\extrarowheight{3pt}
\scalebox{0.95}{
\begin{tabular}{|c||c|c|c|}
\hline 
  \multicolumn{4}{|c|}{Pull$_{\rm SM}$ with $R_{K^{(*)}}$ and ${\rm BR}(B_s\to \mu^+ \mu^-)$ prospects} \\ 
\hline 
 LHCb lum.		 &  18 fb$^{-1}$ & 50 fb$^{-1}$   &   300 fb$^{-1}$  \\
\hline
\hline
$\delta C_{9}^{\mu}$	 	 & $ 6.5\sigma	 $ & $ 14.7\sigma	 $ &  $ 21.9\sigma	 $   \\
$\delta C_{10}^{\mu}$	 	 & $ 7.1\sigma	 $ & $ 16.6\sigma	 $ &  $ 25.1\sigma	 $   \\
$\delta C_{LL}^{\mu}$	 	 & $ 7.5\sigma	 $ & $ 17.7\sigma	 $ &  $ 26.6\sigma	 $   \\
\hline
\end{tabular}
}\vspace*{0.1cm}
\caption{\small
Predictions of Pull$_{\rm SM}$ for the fit to $\delta C_9^\mu$, $\delta C_{10}^\mu$ and $\delta C_{LL}^\mu$ (as given in the right panel of Table~\ref{tab:ComparisonRKRKstarBmumu_1DA}) for the LHCb upgrade scenarios with 18, 50 and 300 fb$^{-1}$ luminosity collected.
\label{tab:upgrade+Bsmumu}}
\end{center} 
\end{table}

However, it should be noted that the significance is rather strongly dependent on the presumed systematic uncertainties as well as the considered scenario.  
This can be seen in Fig.~\ref{fig:projections} where Pull$_{\rm SM}$ for each LFUV ratios is individually shown for the $C_{9}^\mu$ and $C_{10}^\mu$ scenarios with different assumptions on the systematic uncertainties. 
For the $C_{9}^\mu$ case, $R_K$ can individually reach $5\sigma$ significance at $\sim16$ fb$^{-1}$ luminosity. 
On the one hand for the $C_{10}^\mu$ scenario, with the ultimate systematic uncertainty for $R_{K^*}([1.1,6])$ it gives $5\sigma$ Pull$_{\rm SM}$ at $\sim13$ fb$^{-1}$, however, assuming the current systematic uncertainty remains, it does not reach $3\sigma$ significance. On the other hand for the same  $C_{10}^\mu$ scenario, with $R_K$, there is $5\sigma$ significance with $\sim20$ fb$^{-1}$ luminosity and it is much less dependent on the assumption on the systematic error. 
In both scenarios $R_{K^*}([0.045,1.1])$ does not give a large Pull$_{\rm SM}$
as it is mostly dominated by $C_7$ and with the considered NP scenarios the predicted value for this bin (as given in the next paragraph) are very close to the SM prediction, $0.906 \pm 0.028$.

\begin{figure}[t!]
\begin{center}
\includegraphics[width=0.49\textwidth]{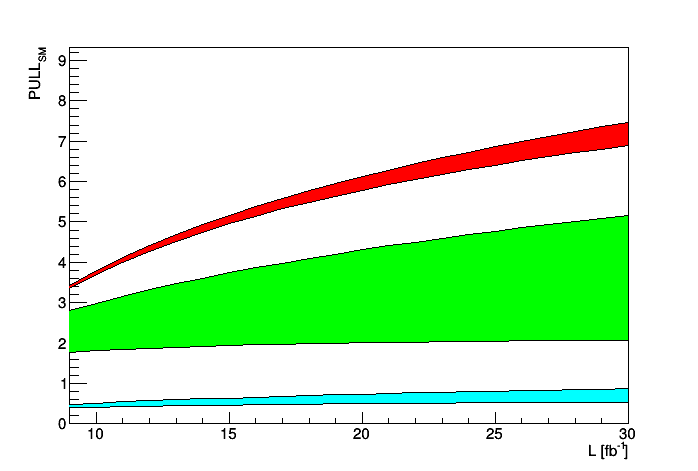}
\includegraphics[width=0.49\textwidth]{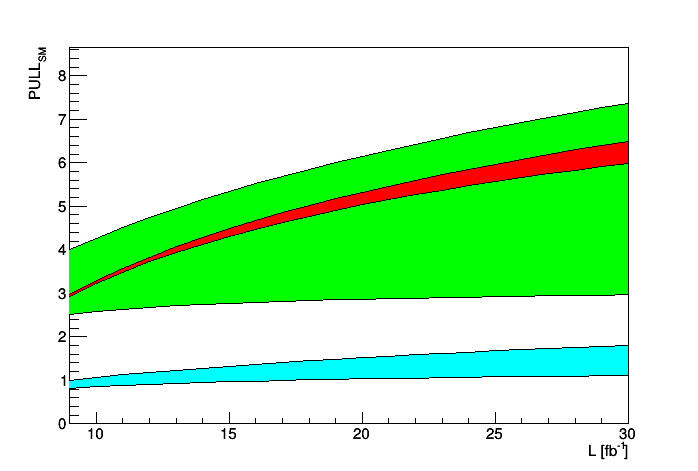}
\caption{\small
Significance of the tension between
SM predictions and the experimental projections for LFUV observables, on the left [right] assuming current central value of $C_9^\mu$ [$C_{10}^\mu$] from the clean observables (right panel of Table~\ref{tab:ComparisonRKRKstarBmumu_1DA}) remains unchanged.
The red, green and blue band correspond to $R_K^{[1.1,6]}$, $R_{K^*}^{[1.1,6]}$ and $R_{K^*}^{[0.045,1.1]}$, respectively. The lower [upper] limit of each band corresponds to assuming the current systematic uncertainties do not improve [ultimate systematic uncertainty  envisaged for the 50 fb$^{-1}$  and 300 fb$^{-1}$ luminosity] (see text for further details). 
\label{fig:projections}}
\end{center}
\end{figure}

Nonetheless, a single LFUV observable cannot individually pinpoint the correct NP scenario.  
For example, the predicted 68\% confidence interval with 9~fb$^{-1}$ for the (very) low bin of $R_{K^{*}}$ within the  $C_{9}^\mu$  scenario is given by ($[0.897,0.874]$) $[0.853,0.906]$ while for the $C_{10}^\mu$ scenario it is ($[0.871,0.889]$) $[0.793,0.865]$ having overlapping intervals which prevents disentangling  the preferred scenario. Considering the other four scenarios of Table~\ref{tab:ComparisonRKRKstarBmumu_1DA} this problem becomes even more pronounced. The analytical dependence of the ratios on the NP-WCs given in Ref.~\cite{DAmico:2017mtc}  explain this feature. So it is expected that this feature stays valid also in future scenarios. 

In Table~\ref{tab:RxPredictions9fb} we give the 68\% confidence interval predictions for other LFUV ratios of muons in the final state over electrons, assuming the various NP fits to $R_{K^{(*)}}$ and BR($B_s \to \mu^+ \mu^-$) as given in the right panel of Table~\ref{tab:ComparisonRKRKstarBmumu_1DA}. 
There are a number of the ratios which are able to discern among the various scenarios (see also~\cite{Hurth:2017hxg}).
From the first row of Table~\ref{tab:RxPredictions9fb}, $R_{F_L}([1.1,6])$ is predicted to have distinct intervals whether the considered scenario is NP in $C_{9}^\mu$ or $C_{10}^\mu$, however there are still overlaps with other cases e.g. $C_{9}^e$ which can be disentangled to some extent by considering further observables such as $R_{S_5}([1.1,6])$. 

In Table~\ref{tab:RxPredictions18fb} and~\ref{tab:RxPredictions50fb} we give the 1$\sigma$ range predictions of these LFUV observables for the 18 and 50 fb$^{-1}$ luminosity benchmarks, respectively where several of the observables give more distinct predictions for the various NP scenarios.  

In Table~\ref{tab:RxPredictions9fb}, some of the observables such as $R_{A{{\rm FB}}}([1.1,6])$ have a rather large uncertainty which is due to zero-crossings. 
In such cases, it is more suitable to consider observable differences ($O_i^\mu - O_i^e$) instead of ratios $O_i^\mu/O_i^e$~\cite{Altmannshofer:2015mqa}. Similar observables are also defined for the optimised $P_i^{(\prime)}$ observables in~\cite{Capdevila:2016ivx}. 
In Appendix~\ref{app:AltObs}, we give the prediction for the alternative set of observables in Tables~\ref{tab:RestPredictions9fb},~\ref{tab:RestPredictions18fb} and~\ref{tab:RestPredictions50fb} for 9, 18 and 50 fb$^{-1}$ luminosities, respectively.

\begin{table}[h!]
\begin{center}
\rowcolors{3}{}{light-gray}
\setlength\extrarowheight{3pt}
\scalebox{0.85}{
\begin{tabular}{|l||c|c|c|c|c|c|}
\hline 
 & \multicolumn{6}{c|}{Predictions with best fit values of ``clean'' observables} \\ 
\hline
Obs. & $C_{9}^{\mu}$ & $C_{9}^{e}$ & $C_{10}^{\mu}$ & $C_{10}^{e}$ & $C_{LL}^{\mu}$& $C_{LL}^{e}$ \\																									
\hline																									
$R_{F_L}^{[1.1,6.0]}$ 	& $ [	0.910	,	0.943	]$ &  $[	0.935	,	0.951	]$ &  $[	0.992	,	1.001	]$ &  $[	0.993	,	1.000	]$ &  $[	0.957	,	0.968	]$ &  $[	1.000	,	1.016	] $ \\
$R_{A_{FB}}^{[1.1,6.0]}$	& $ [	3.909	,	6.457	]$ &  $[	-0.572	,	-0.296	]$ &  $[	0.937	,	0.939	]$ &  $[	0.953	,	0.980	]$ &  $[	2.443	,	3.488	]$ &  $[	0.244	,	0.361	] $ \\
$R_{S_{3}}^{[1.1,6.0]}$	& $ [	0.912	,	0.942	]$ &  $[	0.900	,	0.936	]$ &  $[	0.803	,	0.880	]$ &  $[	0.841	,	0.891	]$ &  $[	0.826	,	0.895	]$ &  $[	1.020	,	1.040	] $ \\
$R_{S_{5}}^{[1.1,6.0]}$	& $ [	0.335	,	0.650	]$ &  $[	0.697	,	0.778	]$ &  $[	1.013	,	1.014	]$ &  $[	1.030	,	1.059	]$ &  $[	0.731	,	0.840	]$ &  $[	1.213	,	1.475	] $ \\
$R_{F_L}^{[15,19]}$	& $ [	0.998	,	0.999	]$ &  $[	0.998	,	0.998	]$ &  $[	0.998	,	0.998	]$ &  $[	0.998	,	0.998	]$ &  $[	0.998	,	0.998	]$ &  $[	0.998	,	0.998	] $ \\
$R_{A_{FB}}^{[15,19]}$	& $ [	0.908	,	0.961	]$ &  $[	0.988	,	0.992	]$ &  $[	1.007	,	1.010	]$ &  $[	1.027	,	1.053	]$ &  $[	0.995	,	0.997	]$ &  $[	1.017	,	1.036	] $ \\
$R_{S_{3}}^{[15,19]}$	& $ [	0.998	,	0.998	]$ &  $[	0.998	,	0.998	]$ &  $[	0.999	,	0.999	]$ &  $[	0.999	,	0.999	]$ &  $[	0.999	,	0.999	]$ &  $[	0.998	,	0.998	] $ \\
$R_{S_{5}}^{[15,19]}$	& $ [	0.908	,	0.960	]$ &  $[	0.988	,	0.992	]$ &  $[	1.007	,	1.010	]$ &  $[	1.027	,	1.053	]$ &  $[	0.995	,	0.997	]$ &  $[	1.017	,	1.036	] $ \\
$R_{K^*}^{[15,19]}$	& $ [	0.798	,	0.877	]$ &  $[	0.785	,	0.868	]$ &  $[	0.803	,	0.873	]$ &  $[	0.768	,	0.853	]$ &  $[	0.773	,	0.856	]$ &  $[	1.072	,	1.128	] $ \\
$R_{K}^{[15,19]}$	& $ [	0.795	,	0.877	]$ &  $[	0.789	,	0.871	]$ &  $[	0.833	,	0.893	]$ &  $[	0.796	,	0.874	]$ &  $[	0.791	,	0.869	]$ &  $[	1.077	,	1.135	] $ \\
$R_{\phi}^{[1.1,6.0]}$	& $ [	0.843	,	0.901	]$ &  $[	0.813	,	0.887	]$ &  $[	0.799	,	0.869	]$ &  $[	0.764	,	0.850	]$ &  $[	0.792	,	0.867	]$ &  $[	1.054	,	1.096	] $ \\
$R_{\phi}^{[15,19]}$	& $ [	0.799	,	0.877	]$ &  $[	0.785	,	0.868	]$ &  $[	0.801	,	0.872	]$ &  $[	0.766	,	0.853	]$ &  $[	0.772	,	0.856	]$ &  $[	1.072	,	1.128	] $ \\
\hline																									
\end{tabular}
}
\caption{\small
Predictions of LFUV observables at 68\% confidence level, considering one operator fits obtained with the clean observables of Table~\ref{tab:ComparisonRKRKstarBmumu_1DA}.
The observables $R_{K^{(*)}}$ and $R_\phi$ refer to the the branching fraction ratios of $B\to K^{(*)} \bar \ell \ell$ and $B_s\to \phi \bar \ell \ell$, respectively.
The other observables correspond to ratios of the angular observables of the $B\to K^{*} \bar \ell \ell$ decay and the superscripts denote the $q^2$ bins.
\label{tab:RxPredictions9fb}} 
\end{center} 
\end{table}
\begin{table}[h!]
\begin{center}
\rowcolors{3}{}{light-gray}
\setlength\extrarowheight{3pt}
\scalebox{0.85}{
\begin{tabular}{|l||c|c|c|c|c|c|}
\hline 
 & \multicolumn{6}{c|}{Predictions assuming 18 fb$^{-1}$ luminosity} \\ 
\hline
Obs. & $C_{9}^{\mu}$ & $C_{9}^{e}$ & $C_{10}^{\mu}$ & $C_{10}^{e}$ & $C_{LL}^{\mu}$& $C_{LL}^{e}$ \\																									
\hline																									
$R_{F_L}^{[1.1,6.0]}$ 	& $ [	0.916	,	0.937	]$ &  $[	0.938	,	0.947	]$ &  $[	0.993	,	1.000	]$ &  $[	0.994	,	0.999	]$ &  $[	0.959	,	0.966	]$ &  $[	1.003	,	1.013	] $ \\
$R_{A_{FB}}^{[1.1,6.0]}$	& $ [	4.375	,	5.954	]$ &  $[	-0.480	,	-0.323	]$ &  $[	0.938	,	0.939	]$ &  $[	0.958	,	0.975	]$ &  $[	2.611	,	3.308	]$ &  $[	0.259	,	0.333	] $ \\
$R_{S_{3}}^{[1.1,6.0]}$	& $ [	0.917	,	0.936	]$ &  $[	0.907	,	0.929	]$ &  $[	0.813	,	0.870	]$ &  $[	0.850	,	0.880	]$ &  $[	0.838	,	0.884	]$ &  $[	1.023	,	1.036	] $ \\
$R_{S_{5}}^{[1.1,6.0]}$	& $ [	0.398	,	0.593	]$ &  $[	0.710	,	0.759	]$ &  $[	1.013	,	1.014	]$ &  $[	1.035	,	1.053	]$ &  $[	0.750	,	0.823	]$ &  $[	1.252	,	1.417	] $ \\
$R_{F_L}^{[15,19]}$	& $ [	0.998	,	0.999	]$ &  $[	0.998	,	0.998	]$ &  $[	0.998	,	0.998	]$ &  $[	0.998	,	0.998	]$ &  $[	0.998	,	0.998	]$ &  $[	0.998	,	0.998	] $ \\
$R_{A_{FB}}^{[15,19]}$	& $ [	0.920	,	0.952	]$ &  $[	0.988	,	0.991	]$ &  $[	1.008	,	1.010	]$ &  $[	1.032	,	1.047	]$ &  $[	0.995	,	0.996	]$ &  $[	1.020	,	1.032	] $ \\
$R_{S_{3}}^{[15,19]}$	& $ [	0.998	,	0.998	]$ &  $[	0.998	,	0.998	]$ &  $[	0.999	,	0.999	]$ &  $[	0.999	,	0.999	]$ &  $[	0.999	,	0.999	]$ &  $[	0.998	,	0.998	] $ \\
$R_{S_{5}}^{[15,19]}$	& $ [	0.919	,	0.952	]$ &  $[	0.988	,	0.990	]$ &  $[	1.008	,	1.010	]$ &  $[	1.032	,	1.047	]$ &  $[	0.995	,	0.996	]$ &  $[	1.020	,	1.032	] $ \\
$R_{K^*}^{[15,19]}$	& $ [	0.812	,	0.861	]$ &  $[	0.800	,	0.851	]$ &  $[	0.812	,	0.863	]$ &  $[	0.784	,	0.835	]$ &  $[	0.787	,	0.842	]$ &  $[	1.082	,	1.118	] $ \\
$R_{K}^{[15,19]}$	& $ [	0.810	,	0.861	]$ &  $[	0.804	,	0.854	]$ &  $[	0.841	,	0.885	]$ &  $[	0.811	,	0.858	]$ &  $[	0.804	,	0.856	]$ &  $[	1.087	,	1.124	] $ \\
$R_{\phi}^{[1.1,6.0]}$	& $ [	0.853	,	0.889	]$ &  $[	0.827	,	0.873	]$ &  $[	0.808	,	0.859	]$ &  $[	0.780	,	0.832	]$ &  $[	0.804	,	0.854	]$ &  $[	1.062	,	1.088	] $ \\
$R_{\phi}^{[15,19]}$	& $ [	0.813	,	0.861	]$ &  $[	0.800	,	0.851	]$ &  $[	0.810	,	0.862	]$ &  $[	0.783	,	0.834	]$ &  $[	0.786	,	0.842	]$ &  $[	1.082	,	1.118	] $ \\
\hline																									
\end{tabular}
}
\caption{\small
Predictions of ratios at 68\% confidence level for different scenarios, assuming the  central values of Table~\ref{tab:ComparisonRKRKstarBmumu_1DA} remains the same with 18 fb$^{-1}$ luminosity.
See the caption of Table~\ref{tab:RxPredictions9fb} for more details.
\label{tab:RxPredictions18fb}} 
\end{center} 
\end{table}
\begin{table}[h!]
\begin{center}
\rowcolors{3}{}{light-gray}
\setlength\extrarowheight{3pt}
\scalebox{0.85}{
\begin{tabular}{|l||c|c|c|c|c|c|}
\hline 
 & \multicolumn{6}{c|}{Predictions assuming 50 fb$^{-1}$ luminosity} \\ 
\hline
Obs. & $C_{9}^{\mu}$ & $C_{9}^{e}$ & $C_{10}^{\mu}$ & $C_{10}^{e}$ & $C_{LL}^{\mu}$& $C_{LL}^{e}$ \\																									
\hline																									
$R_{F_L}^{[1.1,6.0]}$ 	& $ [	0.922	,	0.932	]$ &  $[	0.941	,	0.944	]$ &  $[	0.995	,	0.998	]$ &  $[	0.996	,	0.997	]$ &  $[	0.961	,	0.964	]$ &  $[	1.006	,	1.010	] $ \\
$R_{A_{FB}}^{[1.1,6.0]}$	& $ [	4.791	,	5.520	]$ &  $[	-0.416	,	-0.358	]$ &  $[	0.938	,	0.939	]$ &  $[	0.963	,	0.970	]$ &  $[	2.822	,	3.089	]$ &  $[	0.279	,	0.307	] $ \\
$R_{S_{3}}^{[1.1,6.0]}$	& $ [	0.922	,	0.931	]$ &  $[	0.914	,	0.922	]$ &  $[	0.832	,	0.852	]$ &  $[	0.858	,	0.870	]$ &  $[	0.853	,	0.870	]$ &  $[	1.027	,	1.032	] $ \\
$R_{S_{5}}^{[1.1,6.0]}$	& $ [	0.453	,	0.543	]$ &  $[	0.723	,	0.742	]$ &  $[	1.014	,	1.014	]$ &  $[	1.040	,	1.048	]$ &  $[	0.773	,	0.801	]$ &  $[	1.298	,	1.361	] $ \\
$R_{F_L}^{[15,19]}$	& $ [	0.998	,	0.999	]$ &  $[	0.998	,	0.998	]$ &  $[	0.998	,	0.998	]$ &  $[	0.998	,	0.998	]$ &  $[	0.998	,	0.998	]$ &  $[	0.998	,	0.998	] $ \\
$R_{A_{FB}}^{[15,19]}$	& $ [	0.929	,	0.944	]$ &  $[	0.988	,	0.989	]$ &  $[	1.009	,	1.010	]$ &  $[	1.036	,	1.042	]$ &  $[	0.996	,	0.996	]$ &  $[	1.023	,	1.028	] $ \\
$R_{S_{3}}^{[15,19]}$	& $ [	0.998	,	0.998	]$ &  $[	0.998	,	0.998	]$ &  $[	0.999	,	0.999	]$ &  $[	0.999	,	0.999	]$ &  $[	0.999	,	0.999	]$ &  $[	0.998	,	0.998	] $ \\
$R_{S_{5}}^{[15,19]}$	& $ [	0.929	,	0.944	]$ &  $[	0.988	,	0.989	]$ &  $[	1.009	,	1.010	]$ &  $[	1.036	,	1.042	]$ &  $[	0.996	,	0.996	]$ &  $[	1.023	,	1.028	] $ \\
$R_{K^*}^{[15,19]}$	& $ [	0.825	,	0.847	]$ &  $[	0.815	,	0.835	]$ &  $[	0.828	,	0.846	]$ &  $[	0.799	,	0.820	]$ &  $[	0.804	,	0.825	]$ &  $[	1.093	,	1.107	] $ \\
$R_{K}^{[15,19]}$	& $ [	0.823	,	0.847	]$ &  $[	0.819	,	0.838	]$ &  $[	0.854	,	0.870	]$ &  $[	0.825	,	0.844	]$ &  $[	0.820	,	0.839	]$ &  $[	1.098	,	1.113	] $ \\
$R_{\phi}^{[1.1,6.0]}$	& $ [	0.862	,	0.879	]$ &  $[	0.841	,	0.858	]$ &  $[	0.824	,	0.843	]$ &  $[	0.795	,	0.816	]$ &  $[	0.819	,	0.839	]$ &  $[	1.070	,	1.080	] $ \\
$R_{\phi}^{[15,19]}$	& $ [	0.825	,	0.847	]$ &  $[	0.815	,	0.835	]$ &  $[	0.826	,	0.845	]$ &  $[	0.797	,	0.819	]$ &  $[	0.803	,	0.824	]$ &  $[	1.093	,	1.107	] $ \\
\hline																									
\end{tabular}
}
\caption{\small
Predictions of ratios 
at 68\% confidence level for different scenarios, assuming  the central values of Table~\ref{tab:ComparisonRKRKstarBmumu_1DA} remains the same with 50 fb$^{-1}$ luminosity.
See the caption of Table~\ref{tab:RxPredictions9fb} for more details.
\label{tab:RxPredictions50fb}} 
\end{center} 
\end{table}

\section{Conclusions}\label{sec:conclusions}
The current experimental data on $b \to s$ transitions show deviations in several observables with respect to  the Standard Model predictions. 
The latest LHCb update of the leptonic decay BR($B_s \to \mu^+ \mu^-$) and the lepton flavour violating ratio $R_K$ have further strengthened the New Physics description of the so-called $B$-anomalies which when taken together with the previously measured anomalies in the two bins of $R_{K^*}$ results in more than 4$\sigma$ significance.

Considering all available observables of $b \to s \ell \ell$ processes, the significance of the improved description of the data by New Physics contributions becomes even higher, this is however reliant on the assumptions made on the size of the not well-known power corrections in a number of observables of the exclusive $B \to K^{(*)}\ell \ell$ and $B_s \to \phi \ell \ell$ decays. In order to have an unbiased determination of the structure of New Physics contributions, we considered a 20-dimensional fit where all the relevant $b \to s$ operators are taken in account, still finding a large Pull$_{\rm SM}$. Interestingly, while in our previous analysis there was no significant indication of preference for going beyond one or two operator fits, we now see such an indication for considering simultaneously electron and muon contributions.  

Assuming any of the favoured NP descriptions of the lepton-flavour universality violating observables remain, we show that New Physics can be established with more than $5\sigma$ significance already with 18~fb$^{-1}$ of integrated luminosity. However, the preferred scenario, in general cannot be determined by only considering the $R_{K^{(*)}}$ observables. 
To disentangle the preferred New Physics scenario we also give predictions for further ratios with data already gathered by LHCb as well as the projected data with 18 and 50 fb$^{-1}$ luminosity.

\section*{Acknowledgement}
The work was supported by  the  Cluster  of  Excellence  ``Precision  Physics,  Fundamental
Interactions, and Structure of Matter" (PRISMA$^+$ EXC 2118/1) funded by the German Research Foundation (DFG) within the German Excellence Strategy (Project ID 39083149), as well as BMBF Verbundprojekt 05H2018 - Belle II.  
TH thanks the CERN theory group for its hospitality during his regular visits to CERN where part of the work was done. This work has received financial support from Xunta de Galicia (Centro singular de investigaci\'on de Galicia accreditation 2019-2022), by European Union ERDF, and by  the ''Mar\'ia de Maeztu`` 
Units of Excellence program MDM-2016-0692 and the Spanish Research State Agency. 

\clearpage

\appendix

\section{Predictions for further LFUV observables}\label{app:AltObs}
\begin{table}[h!]
\begin{center}
\vspace*{1.cm}
\rowcolors{3}{}{light-gray}
\setlength\extrarowheight{3pt}
\scalebox{0.8}{
\begin{tabular}{|l||c|c|c|c|c|c|}
\hline 
 & \multicolumn{6}{c|}{Predictions with best fit values of ``clean'' observables} \\ 
\hline
Obs. & $C_{9}^{\mu}$ & $C_{9}^{e}$ & $C_{10}^{\mu}$ & $C_{10}^{e}$ & $C_{LL}^{\mu}$& $C_{LL}^{e}$ \\																									
\hline																									
$D_{F_L}^{[1.1,6.0]}$ 	& $ [	-0.069	,	-0.043	]$ &  $[	-0.052	,	-0.039	]$ &  $[	-0.006	,	0.001	]$ &  $[	-0.005	,	0.000	]$ &  $[	-0.032	,	-0.025	]$ &  $[	0.000	,	0.012	] $ \\
$D_{A_{FB}}^{[1.1,6.0]}$ 	& $ [	-0.074	,	-0.039	]$ &  $[	-0.055	,	-0.034	]$ &  $[	0.001	,	0.001	]$ &  $[	0.000	,	0.001	]$ &  $[	-0.034	,	-0.019	]$ &  $[	0.022	,	0.039	] $ \\
$D_{S_{3}}^{[1.1,6.0]}$ 	& $ [	0.001	,	0.001	]$ &  $[	0.001	,	0.001	]$ &  $[	0.001	,	0.002	]$ &  $[	0.001	,	0.002	]$ &  $[	0.001	,	0.002	]$ &  $[	0.000	,	0.000	] $ \\
$D_{S_{4}}^{[1.1,6.0]}$ 	& $ [	-0.003	,	-0.003	]$ &  $[	-0.009	,	-0.005	]$ &  $[	-0.028	,	-0.017	]$ &  $[	-0.027	,	-0.018	]$ &  $[	-0.020	,	-0.012	]$ &  $[	0.000	,	0.000	] $ \\
$D_{S_{5}}^{[1.1,6.0]}$ 	& $ [	0.057	,	0.108	]$ &  $[	0.046	,	0.071	]$ &  $[	-0.002	,	-0.002	]$ &  $[	-0.009	,	-0.005	]$ &  $[	0.026	,	0.044	]$ &  $[	-0.052	,	-0.028	] $ \\
$D_{F_L}^{[15,19]}$ 	& $ [	-0.001	,	0.000	]$ &  $[	-0.001	,	-0.001	]$ &  $[	-0.001	,	-0.001	]$ &  $[	-0.001	,	-0.001	]$ &  $[	-0.001	,	-0.001	]$ &  $[	-0.001	,	-0.001	] $ \\
$D_{A_{FB}}^{[15,19]}$ 	& $ [	-0.035	,	-0.015	]$ &  $[	-0.005	,	-0.003	]$ &  $[	0.003	,	0.004	]$ &  $[	0.010	,	0.019	]$ &  $[	-0.002	,	-0.001	]$ &  $[	0.006	,	0.013	] $ \\
$D_{S_{3}}^{[15,19]}$ 	& $ [	0.000	,	0.000	]$ &  $[	0.000	,	0.000	]$ &  $[	0.000	,	0.000	]$ &  $[	0.000	,	0.000	]$ &  $[	0.000	,	0.000	]$ &  $[	0.000	,	0.000	] $ \\
$D_{S_{4}}^{[15,19]}$ 	& $ [	0.000	,	0.000	]$ &  $[	-0.001	,	-0.001	]$ &  $[	-0.001	,	-0.001	]$ &  $[	-0.001	,	0.000	]$ &  $[	-0.001	,	-0.001	]$ &  $[	-0.001	,	-0.001	] $ \\
$D_{S_{5}}^{[15,19]}$ 	& $ [	0.011	,	0.026	]$ &  $[	0.002	,	0.004	]$ &  $[	-0.003	,	-0.002	]$ &  $[	-0.014	,	-0.007	]$ &  $[	0.001	,	0.001	]$ &  $[	-0.010	,	-0.005	] $ \\
\hline																									
$R_{P_{2}}^{[1.1,6.0]}$ 	& $ [	3.557	,	5.358	]$ &  $[	-0.526	,	-0.255	]$ &  $[	0.998	,	1.030	]$ &  $[	1.012	,	1.063	]$ &  $[	2.397	,	3.320	]$ &  $[	0.277	,	0.392	] $ \\
$R_{P_{1}}^{[1.1,6.0]}$ 	& $ [	0.757	,	0.858	]$ &  $[	0.778	,	0.000	]$ &  $[	0.881	,	0.936	]$ &  $[	0.912	,	0.945	]$ &  $[	0.786	,	0.878	]$ &  $[	1.105	,	1.180	] $ \\
$R_{P_{4}}^{[1.1,6.0]}$ 	& $ [	0.934	,	0.959	]$ &  $[	0.902	,	-0.039	]$ &  $[	0.821	,	0.899	]$ &  $[	0.862	,	0.910	]$ &  $[	0.845	,	0.912	]$ &  $[	1.039	,	1.059	] $ \\
$R_{P_{5}}^{[1.1,6.0]}$ 	& $ [	0.320	,	0.638	]$ &  $[	0.670	,	-0.034	]$ &  $[	1.049	,	1.062	]$ &  $[	1.065	,	1.103	]$ &  $[	0.729	,	0.846	]$ &  $[	1.263	,	1.559	] $ \\
$R_{P_{2}}^{[15,19]}$ 	& $ [	0.909	,	0.962	]$ &  $[	0.990	,	0.001	]$ &  $[	1.010	,	1.012	]$ &  $[	1.029	,	1.055	]$ &  $[	0.997	,	0.999	]$ &  $[	1.019	,	1.038	] $ \\
$R_{P_{1}}^{[15,19]}$ 	& $ [	1.000	,	1.000	]$ &  $[	1.000	,	-0.005	]$ &  $[	1.001	,	1.001	]$ &  $[	1.001	,	1.001	]$ &  $[	1.000	,	1.001	]$ &  $[	1.000	,	1.000	] $ \\
$R_{P_{4}}^{[15,19]}$ 	& $ [	1.000	,	1.000	]$ &  $[	1.000	,	0.071	]$ &  $[	1.000	,	1.000	]$ &  $[	1.000	,	1.000	]$ &  $[	1.000	,	1.000	]$ &  $[	1.000	,	1.000	] $ \\
$R_{P_{5}}^{[15,19]}$ 	& $ [	0.909	,	0.962	]$ &  $[	0.989	,	-0.001	]$ &  $[	1.010	,	1.012	]$ &  $[	1.029	,	1.055	]$ &  $[	0.997	,	0.999	]$ &  $[	1.019	,	1.038	] $ \\
\hline																									
$Q_{2}^{[1.1,6.0]}$ 	& $ [	0.096	,	0.164	]$ &  $[	0.103	,	0.174	]$ &  $[	0.000	,	0.001	]$ &  $[	0.000	,	0.002	]$ &  $[	0.053	,	0.087	]$ &  $[	-0.093	,	-0.055	] $ \\
$Q_{1}^{[1.1,6.0]}$ 	& $ [	0.014	,	0.024	]$ &  $[	0.016	,	0.029	]$ &  $[	0.006	,	0.012	]$ &  $[	0.006	,	0.010	]$ &  $[	0.012	,	0.021	]$ &  $[	-0.015	,	-0.010	] $ \\
$Q_{4}^{[1.1,6.0]}$ 	& $ [	-0.041	,	-0.025	]$ &  $[	-0.068	,	-0.036	]$ &  $[	-0.110	,	-0.062	]$ &  $[	-0.100	,	-0.061	]$ &  $[	-0.095	,	-0.054	]$ &  $[	0.023	,	0.035	] $ \\
$Q_{5}^{[1.1,6.0]}$ 	& $ [	0.138	,	0.259	]$ &  $[	0.119	,	0.191	]$ &  $[	-0.024	,	-0.019	]$ &  $[	-0.036	,	-0.024	]$ &  $[	0.059	,	0.104	]$ &  $[	-0.139	,	-0.081	] $ \\
$Q_{2}^{[15,19]}$ 	& $ [	0.014	,	0.034	]$ &  $[	0.002	,	0.004	]$ &  $[	-0.004	,	-0.004	]$ &  $[	-0.020	,	-0.011	]$ &  $[	0.001	,	0.001	]$ &  $[	-0.014	,	-0.007	] $ \\
$Q_{1}^{[15,19]}$ 	& $ [	0.000	,	0.000	]$ &  $[	0.000	,	0.000	]$ &  $[	-0.001	,	0.000	]$ &  $[	-0.001	,	0.000	]$ &  $[	0.000	,	0.000	]$ &  $[	0.000	,	0.000	] $ \\
$Q_{4}^{[15,19]}$ 	& $ [	0.000	,	0.000	]$ &  $[	0.000	,	0.000	]$ &  $[	0.000	,	0.000	]$ &  $[	0.000	,	0.000	]$ &  $[	0.000	,	0.000	]$ &  $[	0.000	,	0.000	] $ \\
$Q_{5}^{[15,19]}$ 	& $ [	0.023	,	0.055	]$ &  $[	0.004	,	0.006	]$ &  $[	-0.007	,	-0.006	]$ &  $[	-0.031	,	-0.017	]$ &  $[	0.001	,	0.002	]$ &  $[	-0.022	,	-0.011	] $ \\
\hline																									
\end{tabular}
}
\caption{\small
Predictions of ratios and differences of observables with muons in the final state to electrons in the final state at 68\% confidence level, considering one operator fits obtained with the clean observables of Table~\ref{tab:ComparisonRKRKstarBmumu_1DA}.
The observables $D_{S_{3,4,5}} = S_{3,4,5}^{\mu}-S_{3,4,5}^{e}$, $D_{A_{FB}} = A_{FB}^{\mu}-A_{FB}^{e}$, $D_{F_L} = Q_{F_L} = F_{L}^{\mu}-F_{L}^{e}$ 
and $Q_{1,2,4,5} = P_{1,2,4,5}^{(\prime)\;\mu}-P_{1,2,4,5}^{(\prime)\;e}$ and $R_{P_i} = P_{1,2,4,5}^{(\prime)\;\mu} /P_{1,2,4,5}^{(\prime)\;e} $
all correspond to the  $B\to K^* \bar \ell \ell$ decay. 
The observables $R_{K^{(*)}},R_{X_s}$ and $R_\phi$ correspond to the ratios of the branching fractions of $B\to K^{(*)} \bar \ell \ell,B \to X_s \bar \ell \ell$
and $B_s\to \phi \bar \ell \ell$, respectively. 
The superscripts  denote the $q^2$ bins.
\label{tab:RestPredictions9fb}} 
\end{center} 
\end{table}

\begin{table}
\begin{center}
\rowcolors{3}{}{light-gray}
\setlength\extrarowheight{3pt}
\scalebox{0.8}{
\begin{tabular}{|l||c|c|c|c|c|c|}
\hline 
 & \multicolumn{6}{c|}{Predictions assuming 18 fb$^{-1}$ luminosity} \\ 
\hline
Obs. & $C_{9}^{\mu}$ & $C_{9}^{e}$ & $C_{10}^{\mu}$ & $C_{10}^{e}$ & $C_{LL}^{\mu}$& $C_{LL}^{e}$ \\																									
\hline																									
$D_{F_L}^{[1.1,6.0]}$ 	& $ [	-0.064	,	-0.048	]$ &  $[	-0.049	,	-0.041	]$ &  $[	-0.005	,	0.000	]$ &  $[	-0.004	,	-0.001	]$ &  $[	-0.031	,	-0.026	]$ &  $[	0.002	,	0.010	] $ \\
$D_{A_{FB}}^{[1.1,6.0]}$ 	& $ [	-0.067	,	-0.046	]$ &  $[	-0.051	,	-0.038	]$ &  $[	0.001	,	0.001	]$ &  $[	0.000	,	0.001	]$ &  $[	-0.031	,	-0.022	]$ &  $[	0.025	,	0.036	] $ \\
$D_{S_{3}}^{[1.1,6.0]}$ 	& $ [	0.001	,	0.001	]$ &  $[	0.001	,	0.001	]$ &  $[	0.002	,	0.002	]$ &  $[	0.002	,	0.002	]$ &  $[	0.001	,	0.002	]$ &  $[	0.000	,	0.000	] $ \\
$D_{S_{4}}^{[1.1,6.0]}$ 	& $ [	-0.003	,	-0.003	]$ &  $[	-0.008	,	-0.006	]$ &  $[	-0.027	,	-0.019	]$ &  $[	-0.025	,	-0.020	]$ &  $[	-0.019	,	-0.014	]$ &  $[	0.000	,	0.000	] $ \\
$D_{S_{5}}^{[1.1,6.0]}$ 	& $ [	0.066	,	0.098	]$ &  $[	0.051	,	0.066	]$ &  $[	-0.002	,	-0.002	]$ &  $[	-0.008	,	-0.006	]$ &  $[	0.029	,	0.041	]$ &  $[	-0.048	,	-0.033	] $ \\
$D_{F_L}^{[15,19]}$ 	& $ [	-0.001	,	0.000	]$ &  $[	-0.001	,	-0.001	]$ &  $[	-0.001	,	-0.001	]$ &  $[	-0.001	,	-0.001	]$ &  $[	-0.001	,	-0.001	]$ &  $[	-0.001	,	-0.001	] $ \\
$D_{A_{FB}}^{[15,19]}$ 	& $ [	-0.030	,	-0.018	]$ &  $[	-0.005	,	-0.004	]$ &  $[	0.003	,	0.004	]$ &  $[	0.012	,	0.017	]$ &  $[	-0.002	,	-0.001	]$ &  $[	0.007	,	0.012	] $ \\
$D_{S_{3}}^{[15,19]}$ 	& $ [	0.000	,	0.000	]$ &  $[	0.000	,	0.000	]$ &  $[	0.000	,	0.000	]$ &  $[	0.000	,	0.000	]$ &  $[	0.000	,	0.000	]$ &  $[	0.000	,	0.000	] $ \\
$D_{S_{4}}^{[15,19]}$ 	& $ [	0.000	,	0.000	]$ &  $[	-0.001	,	-0.001	]$ &  $[	-0.001	,	-0.001	]$ &  $[	-0.001	,	0.000	]$ &  $[	-0.001	,	-0.001	]$ &  $[	-0.001	,	-0.001	] $ \\
$D_{S_{5}}^{[15,19]}$ 	& $ [	0.014	,	0.023	]$ &  $[	0.003	,	0.004	]$ &  $[	-0.003	,	-0.002	]$ &  $[	-0.013	,	-0.009	]$ &  $[	0.001	,	0.001	]$ &  $[	-0.009	,	-0.006	] $ \\
\hline																									
$R_{P_{2}}^{[1.1,6.0]}$ 	& $ [	3.914	,	5.029	]$ &  $[	-0.436	,	-0.282	]$ &  $[	1.002	,	1.026	]$ &  $[	1.022	,	1.053	]$ &  $[	2.551	,	3.166	]$ &  $[	0.292	,	0.364	] $ \\
$R_{P_{1}}^{[1.1,6.0]}$ 	& $ [	0.775	,	0.837	]$ &  $[	0.792	,	0.000	]$ &  $[	0.889	,	0.930	]$ &  $[	0.918	,	0.938	]$ &  $[	0.802	,	0.863	]$ &  $[	1.118	,	1.166	] $ \\
$R_{P_{4}}^{[1.1,6.0]}$ 	& $ [	0.937	,	0.953	]$ &  $[	0.910	,	-0.041	]$ &  $[	0.832	,	0.889	]$ &  $[	0.871	,	0.900	]$ &  $[	0.857	,	0.902	]$ &  $[	1.043	,	1.056	] $ \\
$R_{P_{5}}^{[1.1,6.0]}$ 	& $ [	0.383	,	0.580	]$ &  $[	0.685	,	-0.038	]$ &  $[	1.051	,	1.060	]$ &  $[	1.072	,	1.095	]$ &  $[	0.749	,	0.827	]$ &  $[	1.307	,	1.494	] $ \\
$R_{P_{2}}^{[15,19]}$ 	& $ [	0.921	,	0.954	]$ &  $[	0.990	,	0.001	]$ &  $[	1.010	,	1.012	]$ &  $[	1.034	,	1.049	]$ &  $[	0.997	,	0.998	]$ &  $[	1.022	,	1.034	] $ \\
$R_{P_{1}}^{[15,19]}$ 	& $ [	1.000	,	1.000	]$ &  $[	1.000	,	-0.006	]$ &  $[	1.001	,	1.001	]$ &  $[	1.001	,	1.001	]$ &  $[	1.000	,	1.001	]$ &  $[	1.000	,	1.000	] $ \\
$R_{P_{4}}^{[15,19]}$ 	& $ [	1.000	,	1.000	]$ &  $[	1.000	,	0.066	]$ &  $[	1.000	,	1.000	]$ &  $[	1.000	,	1.000	]$ &  $[	1.000	,	1.000	]$ &  $[	1.000	,	1.000	] $ \\
$R_{P_{5}}^{[15,19]}$ 	& $ [	0.921	,	0.954	]$ &  $[	0.990	,	-0.001	]$ &  $[	1.010	,	1.012	]$ &  $[	1.033	,	1.049	]$ &  $[	0.997	,	0.998	]$ &  $[	1.022	,	1.034	] $ \\
\hline																									
$Q_{2}^{[1.1,6.0]}$ 	& $ [	0.110	,	0.152	]$ &  $[	0.117	,	0.161	]$ &  $[	0.000	,	0.001	]$ &  $[	0.001	,	0.002	]$ &  $[	0.058	,	0.082	]$ &  $[	-0.086	,	-0.062	] $ \\
$Q_{1}^{[1.1,6.0]}$ 	& $ [	0.016	,	0.023	]$ &  $[	0.019	,	0.027	]$ &  $[	0.007	,	0.011	]$ &  $[	0.007	,	0.009	]$ &  $[	0.014	,	0.020	]$ &  $[	-0.014	,	-0.011	] $ \\
$Q_{4}^{[1.1,6.0]}$ 	& $ [	-0.039	,	-0.029	]$ &  $[	-0.062	,	-0.042	]$ &  $[	-0.104	,	-0.069	]$ &  $[	-0.092	,	-0.069	]$ &  $[	-0.088	,	-0.061	]$ &  $[	0.026	,	0.033	] $ \\
$Q_{5}^{[1.1,6.0]}$ 	& $ [	0.160	,	0.235	]$ &  $[	0.134	,	0.178	]$ &  $[	-0.023	,	-0.019	]$ &  $[	-0.034	,	-0.026	]$ &  $[	0.066	,	0.096	]$ &  $[	-0.128	,	-0.091	] $ \\
$Q_{2}^{[15,19]}$ 	& $ [	0.017	,	0.030	]$ &  $[	0.003	,	0.004	]$ &  $[	-0.004	,	-0.004	]$ &  $[	-0.018	,	-0.012	]$ &  $[	0.001	,	0.001	]$ &  $[	-0.012	,	-0.008	] $ \\
$Q_{1}^{[15,19]}$ 	& $ [	0.000	,	0.000	]$ &  $[	0.000	,	0.000	]$ &  $[	-0.001	,	-0.001	]$ &  $[	-0.001	,	-0.001	]$ &  $[	0.000	,	0.000	]$ &  $[	0.000	,	0.000	] $ \\
$Q_{4}^{[15,19]}$ 	& $ [	0.000	,	0.000	]$ &  $[	0.000	,	0.000	]$ &  $[	0.000	,	0.000	]$ &  $[	0.000	,	0.000	]$ &  $[	0.000	,	0.000	]$ &  $[	0.000	,	0.000	] $ \\
$Q_{5}^{[15,19]}$ 	& $ [	0.028	,	0.048	]$ &  $[	0.005	,	0.006	]$ &  $[	-0.007	,	-0.006	]$ &  $[	-0.028	,	-0.020	]$ &  $[	0.001	,	0.002	]$ &  $[	-0.020	,	-0.013	] $ \\
\hline																									
\end{tabular}
}
\caption{\small 
Predictions of ratios of observables with muons in the final state to electrons in the final state at 68\% confidence level for different scenarios, assuming the the central values of Table~\ref{tab:ComparisonRKRKstarBmumu_1DA} remains the same with 18 fb$^{-1}$ luminosity.
For the definition of the observables see the caption of Table~\ref{tab:RestPredictions9fb}.
\label{tab:RestPredictions18fb}} 
\end{center} 
\end{table}

\begin{table}
\begin{center}
\rowcolors{3}{}{light-gray}
\setlength\extrarowheight{3pt}
\scalebox{0.8}{
\begin{tabular}{|l||c|c|c|c|c|c|}
\hline 
 & \multicolumn{6}{c|}{Predictions assuming 50 fb$^{-1}$ luminosity} \\ 
\hline
Obs. & $C_{9}^{\mu}$ & $C_{9}^{e}$ & $C_{10}^{\mu}$ & $C_{10}^{e}$ & $C_{LL}^{\mu}$& $C_{LL}^{e}$ \\																									
\hline																									
$D_{F_L}^{[1.1,6.0]}$ 	& $ [	-0.059	,	-0.052	]$ &  $[	-0.047	,	-0.044	]$ &  $[	-0.004	,	-0.002	]$ &  $[	-0.003	,	-0.002	]$ &  $[	-0.029	,	-0.027	]$ &  $[	0.004	,	0.007	] $ \\
$D_{A_{FB}}^{[1.1,6.0]}$ 	& $ [	-0.061	,	-0.051	]$ &  $[	-0.047	,	-0.042	]$ &  $[	0.001	,	0.001	]$ &  $[	0.000	,	0.000	]$ &  $[	-0.028	,	-0.025	]$ &  $[	0.028	,	0.032	] $ \\
$D_{S_{3}}^{[1.1,6.0]}$ 	& $ [	0.001	,	0.001	]$ &  $[	0.001	,	0.001	]$ &  $[	0.002	,	0.002	]$ &  $[	0.002	,	0.002	]$ &  $[	0.002	,	0.002	]$ &  $[	0.000	,	0.000	] $ \\
$D_{S_{4}}^{[1.1,6.0]}$ 	& $ [	-0.003	,	-0.003	]$ &  $[	-0.007	,	-0.006	]$ &  $[	-0.024	,	-0.021	]$ &  $[	-0.024	,	-0.021	]$ &  $[	-0.017	,	-0.015	]$ &  $[	0.000	,	0.000	] $ \\
$D_{S_{5}}^{[1.1,6.0]}$ 	& $ [	0.074	,	0.089	]$ &  $[	0.056	,	0.062	]$ &  $[	-0.002	,	-0.002	]$ &  $[	-0.007	,	-0.006	]$ &  $[	0.032	,	0.037	]$ &  $[	-0.043	,	-0.037	] $ \\
$D_{F_L}^{[15,19]}$ 	& $ [	-0.001	,	0.000	]$ &  $[	-0.001	,	-0.001	]$ &  $[	-0.001	,	-0.001	]$ &  $[	-0.001	,	-0.001	]$ &  $[	-0.001	,	-0.001	]$ &  $[	-0.001	,	-0.001	] $ \\
$D_{A_{FB}}^{[15,19]}$ 	& $ [	-0.027	,	-0.021	]$ &  $[	-0.004	,	-0.004	]$ &  $[	0.003	,	0.004	]$ &  $[	0.013	,	0.015	]$ &  $[	-0.002	,	-0.001	]$ &  $[	0.009	,	0.010	] $ \\
$D_{S_{3}}^{[15,19]}$ 	& $ [	0.000	,	0.000	]$ &  $[	0.000	,	0.000	]$ &  $[	0.000	,	0.000	]$ &  $[	0.000	,	0.000	]$ &  $[	0.000	,	0.000	]$ &  $[	0.000	,	0.000	] $ \\
$D_{S_{4}}^{[15,19]}$ 	& $ [	0.000	,	0.000	]$ &  $[	-0.001	,	-0.001	]$ &  $[	-0.001	,	-0.001	]$ &  $[	-0.001	,	-0.001	]$ &  $[	-0.001	,	-0.001	]$ &  $[	-0.001	,	-0.001	] $ \\
$D_{S_{5}}^{[15,19]}$ 	& $ [	0.016	,	0.020	]$ &  $[	0.003	,	0.003	]$ &  $[	-0.003	,	-0.003	]$ &  $[	-0.012	,	-0.010	]$ &  $[	0.001	,	0.001	]$ &  $[	-0.008	,	-0.006	] $ \\
\hline																									
$R_{P_{2}}^{[1.1,6.0]}$ 	& $ [	4.221	,	4.736	]$ &  $[	-0.373	,	-0.316	]$ &  $[	1.010	,	1.018	]$ &  $[	1.031	,	1.044	]$ &  $[	2.740	,	2.976	]$ &  $[	0.311	,	0.338	] $ \\
$R_{P_{1}}^{[1.1,6.0]}$ 	& $ [	0.791	,	0.820	]$ &  $[	0.807	,	0.000	]$ &  $[	0.902	,	0.917	]$ &  $[	0.924	,	0.932	]$ &  $[	0.821	,	0.845	]$ &  $[	1.133	,	1.151	] $ \\
$R_{P_{4}}^{[1.1,6.0]}$ 	& $ [	0.941	,	0.948	]$ &  $[	0.918	,	-0.044	]$ &  $[	0.850	,	0.871	]$ &  $[	0.879	,	0.891	]$ &  $[	0.871	,	0.888	]$ &  $[	1.047	,	1.052	] $ \\
$R_{P_{5}}^{[1.1,6.0]}$ 	& $ [	0.437	,	0.528	]$ &  $[	0.701	,	-0.042	]$ &  $[	1.054	,	1.058	]$ &  $[	1.079	,	1.088	]$ &  $[	0.774	,	0.804	]$ &  $[	1.359	,	1.430	] $ \\
$R_{P_{2}}^{[15,19]}$ 	& $ [	0.931	,	0.946	]$ &  $[	0.990	,	0.001	]$ &  $[	1.011	,	1.012	]$ &  $[	1.038	,	1.044	]$ &  $[	0.997	,	0.998	]$ &  $[	1.025	,	1.030	] $ \\
$R_{P_{1}}^{[15,19]}$ 	& $ [	1.000	,	1.000	]$ &  $[	1.000	,	-0.006	]$ &  $[	1.001	,	1.001	]$ &  $[	1.001	,	1.001	]$ &  $[	1.000	,	1.001	]$ &  $[	1.000	,	1.000	] $ \\
$R_{P_{4}}^{[15,19]}$ 	& $ [	1.000	,	1.000	]$ &  $[	1.000	,	0.062	]$ &  $[	1.000	,	1.000	]$ &  $[	1.000	,	1.000	]$ &  $[	1.000	,	1.000	]$ &  $[	1.000	,	1.000	] $ \\
$R_{P_{5}}^{[15,19]}$ 	& $ [	0.930	,	0.946	]$ &  $[	0.990	,	-0.001	]$ &  $[	1.011	,	1.012	]$ &  $[	1.038	,	1.044	]$ &  $[	0.997	,	0.998	]$ &  $[	1.025	,	1.030	] $ \\
\hline																									
$Q_{2}^{[1.1,6.0]}$ 	& $ [	0.121	,	0.141	]$ &  $[	0.131	,	0.148	]$ &  $[	0.000	,	0.001	]$ &  $[	0.001	,	0.001	]$ &  $[	0.066	,	0.074	]$ &  $[	-0.079	,	-0.069	] $ \\
$Q_{1}^{[1.1,6.0]}$ 	& $ [	0.018	,	0.021	]$ &  $[	0.021	,	0.024	]$ &  $[	0.008	,	0.010	]$ &  $[	0.007	,	0.008	]$ &  $[	0.016	,	0.018	]$ &  $[	-0.013	,	-0.012	] $ \\
$Q_{4}^{[1.1,6.0]}$ 	& $ [	-0.036	,	-0.032	]$ &  $[	-0.056	,	-0.048	]$ &  $[	-0.092	,	-0.079	]$ &  $[	-0.086	,	-0.076	]$ &  $[	-0.080	,	-0.069	]$ &  $[	0.028	,	0.031	] $ \\
$Q_{5}^{[1.1,6.0]}$ 	& $ [	0.180	,	0.215	]$ &  $[	0.148	,	0.165	]$ &  $[	-0.022	,	-0.021	]$ &  $[	-0.031	,	-0.028	]$ &  $[	0.075	,	0.086	]$ &  $[	-0.117	,	-0.102	] $ \\
$Q_{2}^{[15,19]}$ 	& $ [	0.021	,	0.026	]$ &  $[	0.003	,	0.004	]$ &  $[	-0.004	,	-0.004	]$ &  $[	-0.016	,	-0.014	]$ &  $[	0.001	,	0.001	]$ &  $[	-0.011	,	-0.009	] $ \\
$Q_{1}^{[15,19]}$ 	& $ [	0.000	,	0.000	]$ &  $[	0.000	,	0.000	]$ &  $[	-0.001	,	-0.001	]$ &  $[	-0.001	,	-0.001	]$ &  $[	0.000	,	0.000	]$ &  $[	0.000	,	0.000	] $ \\
$Q_{4}^{[15,19]}$ 	& $ [	0.000	,	0.000	]$ &  $[	0.000	,	0.000	]$ &  $[	0.000	,	0.000	]$ &  $[	0.000	,	0.000	]$ &  $[	0.000	,	0.000	]$ &  $[	0.000	,	0.000	] $ \\
$Q_{5}^{[15,19]}$ 	& $ [	0.033	,	0.042	]$ &  $[	0.005	,	0.006	]$ &  $[	-0.007	,	-0.007	]$ &  $[	-0.026	,	-0.022	]$ &  $[	0.001	,	0.002	]$ &  $[	-0.018	,	-0.015	] $ \\
\hline																									
\end{tabular}
}
\caption{\small
Predictions of ratios of observables with muons in the final state to electrons in the final state at 68\% confidence level for different scenarios, assuming the the central values of Table~\ref{tab:ComparisonRKRKstarBmumu_1DA} remains the same with 50 fb$^{-1}$ luminosity.
For the definition of the observables see the caption of Table~\ref{tab:RestPredictions9fb}.
\label{tab:RestPredictions50fb}} 
\end{center} 
\end{table}

\clearpage

\providecommand{\href}[2]{#2}\begingroup\raggedright\endgroup

\end{document}